\documentclass[preprint,showpacs,preprintnumbers,amsmath,amssymb]{revtex4}

\usepackage{graphicx}
\usepackage{dcolumn}
\usepackage{bm}

\begin{document} 

\title{ Shape and angular distribution of the 4.438-MeV line from proton inelastic scattering off $^{12}$C }


\author{J. Kiener}   Þ
\email{Jurgen.Kiener@csnsm.in2p3.fr} 
\affiliation{Centre de Sciences Nucl\'eaires et de Sciences de la Mati\`ere (CSNSM), Univ
Paris-Sud, CNRS/IN2P3, Universit\'e Paris-Saclay, 91405 Orsay, France}

\date{\today}

\begin{abstract}  The emission of the 4.438-MeV $\gamma$-ray line in proton inelastic scattering   off $^{12}$C has been investigated in detail. For this, two independent descriptions of the correlated scattering and emission process have been used, one for the direct reaction mechanism and the other for the compound-nucleus (CN) component.  Direct reactions were calculated in the framework of the coupled-channels formalism, while the CN component was described as a superposition of separate resonances with definite spin and parity, treated with the angular momentum coupling theory. Both components were incoherently added and compared to a comprehensive data set on measured line shapes and $\gamma$-ray angular distributions in the proton energy range $E_p$ = 5.44 - 25.0 MeV.  In the range  $E_p$ $\sim$ 14 - 25 MeV, good agreement with data was obtained with a dominating direct reaction component. At lower energy,  an important CN component was required to describe line shape and angular distribution data. In general, a good decription of the data could be found for a CN component with spin and parity corresponding to a known nearby resonance. The new calculation was found to improve significantly the agreement with line shape data  in particular in the region of dominating CN resonances compared to previous calculations. Based on these results, predictions for line shapes and $\gamma$-ray angular distributions up to $E_p$ = 100 MeV were made and applied to $\gamma$-ray emissions in solar flares and proton radiotherapy.  \end{abstract}


\maketitle

\section{Introduction}

The 4.438-MeV line of $^{12}$C is probably the most interesting target for line shape studies of proton-induced reactions for several reasons. The  combination of such a relatively high $\gamma$-ray energy with the low mass of the $^{12}$C nucleus,  leads to significant line broadening and energy shift  even in proton reactions. Moreover,  the lifetime of the emitting 2$^+$, 4.439-MeV level of $^{12}$C is short enough (t$_{1/2}$ = 61 fs) for dominant in-flight emission of the $\gamma$ ray in practically all materials.  The emitting level at 4.439 MeV is furthermore the only excited level below the particle emission threshold in $^{12}$C. It results in negligible feeding by higher-lying levels and simplifies considerably the line shape calculations. 

It is an important component in astrophysical sites with accelerated particle populations, where it is one of the strongest prompt nuclear deexcitation lines.  It is for example regularly observed in strong solar flares \cite{Chupp73,Murphy97,SMMcat,Smith03,Gros04,Fermi_flares} and emission of this line from the inner Galaxy due to interactions of low-energy cosmic rays, is one of the primary science objectives of next-generation $\gamma$-ray observatories \cite{Galcenter,eAstrogam}.  Its main production in these sites comes from inelastic scattering of protons and $\alpha$ particles off $^{12}$C and the $^{16}$O(p,p$\alpha$)$^{12}$C and $^{16}$O($\alpha$,2$\alpha$)$^{12}$C reactions and  reactions of accelerated $^{12}$C and $^{16}$O with ambiant H and He.

The interest of line shape analysis to deduce accelerated-particle properties has early been pointed out \cite{Ramaty76}.  First detailed studies were applied to $\gamma$-ray spectra obtained with space-borne moderate-resolution scintillation detectors \cite{SMKS}, and more recently with high-resolution HP-Ge detectors onboard the Rhessi and INTEGRAL satellites \cite{Smith03,Kiener06,Harris07}. In the latter in particular, detailed line shape analysis could be used to pin down composition and directional distribution of the accelerated-particle populations.  Furthermore, there is significant interest in the monitoring of the dose deposition in radiotherapy with particle beams \cite{DP16}. A good example is provided by devices that detect the prompt $\gamma$ rays resulting from proton interactions in human tissue, rising the demand of detailed and reliable nuclear cross sections \cite{Verburg12,Verburg14,Jeya15}. In this context, carbon and oxygen make up more than 3/4 of the human body mass, making the 4.438-MeV line the strongest prompt emission line produced.

A first global, astrophysically motivated approach for the calculation of the 4.438-MeV $\gamma$-ray line profile was presented by Ramaty, Kozlovsky and Lingenfelter \cite{RKL}. It is based on measured differential proton scattering cross sections and a parameterization for the magnetic-substate population of the excited $^{12}$C, reproducing a few measured line shapes and $\gamma$-ray angular distributions, that were available at that time. This approach was further developed with new experiment data, but encountered problems to reproduce simultaneously line shapes and the $\gamma$-ray angular distribution  \cite{MKR}. The agreement with measured data could be improved for proton energies where excitation of the 2$^+$, 4.439-MeV state in $^{12}$C proceeds dominantly through the direct reaction mechanism.  Werntz, Lang and Kim \cite{Werntz90} used nuclear reaction calculations in the coupled-channels formalism, expected to work well in this case, to obtain a complete description of the correlated proton inelastic scattering and subsequent $\gamma$-ray emission process. 

Kiener, de S\'er\'eville and Tatischeff \cite{lshape} adopted both methods to reproduce simultaneously line shapes and $\gamma$-ray angular distributions of an extensive data set, covering a relatively wide angular and energy range in proton irradiations of carbon and oxygen. The data for the 4.438-MeV line in proton inelastic scattering off $^{12}$C above $E_p$ $\sim$15 MeV could be relatively well described by nuclear reaction calculations with the coupled-channels formalism for the direct component, similar to the findings of  Werntz, Lang and Kim \cite{Werntz90}. At lower energies, however, the CN component becomes important and the direct reaction mechanism failed to reproduce the data. There, measured line shapes and $\gamma$-ray angular distributions could be approximately described with the slightly modified approach of \cite{MKR}. 

Since then, new $\gamma$-ray data for proton inelastic scattering off $^{12}$C are available \cite{t2002} which, together with older data \cite{torion}, form a comprehensive data set of line shapes and angular distributions for the 4.438-MeV line in the proton energy range from threshold to $E_p$ = 25 MeV. There are now, in particular, line shape data close to the energy of each CN resonance that shows up as a distinct peak in the  $\gamma$-ray production cross section in proton reactions with $^{12}$C.  With these data, the present paper aims to significantly improve the accuracy of the line-shape calculations, in  particular in the region of dominating CN resonances. For that, a new method has been employed that  improves specifically the description of line shapes and $\gamma$-ray angular distributions in the region of dominating CN resonances, i.e. from threshold to about $E_p$ = 12 MeV.  At higher proton energies, coupled-channels calculations with a deformed potential for nucleon scattering off $^{12}$C \cite{Meigooni85}  reproduced fairly well the measured data. The current work, which includes explicit CN resonances and the direct reaction component in calculations compared to measured data in wide energy and angular ranges, is, to the knowledge of the present author, the most comprehensive study of $\gamma$-ray line shapes ever performed.


In the following section, the formalism of the calculations for the 4.438-MeV line emission in proton inelastic scattering off $^{12}$C will be outlined. The extended parameter search for these calculations to reproduce measured $\gamma$-ray line shapes and angular distributions is described and the results are discussed in section III. Then the interpolation of the calculations to higher proton energies is discussed and finally applications to solar-flare $\gamma$-ray emission and dose monitoring in proton radiotherapy are presented.

\section{Formalism for angular correlations}

The 4.438-MeV line emission results from the deexcitation of the first excited state of $^{12}$C, 2$^+$ at 4.439 MeV with a half-life of 61 fs. This level belongs to the ground-state rotational band and is thus strongly coupled to the 0$^+$ ground state, which results in relatively high cross sections in proton inelastic scattering off $^{12}$C. The second excited state of $^{12}$C, a 0$^+$ state at 7.654 MeV, is already above the $\alpha$-particle emission threshold and has only a very weak  $\gamma$-decay branching,  as generally  for all higher-excited states of $^{12}$C. 

Lets mention for completeness that there are  two 1$^+$ states at 12.71 and 15.11 MeV, that have a moderate $\gamma$-decay branching to the 4.439-MeV state. The inelastic scattering cross sections to these states are, however, much smaller than to the 4.439-MeV state, and cascades from these levels may therefore be safely ignored for the present calculations. The only potential source of complexity is the 4.445-MeV state in $^{11}$B, whose deexcitaton gives rise to a 4.444-MeV $\gamma$-ray line.  Nuclear reaction calculations predict that it may contribute significantly  via the $^{12}$C(p,2p)$^{11}$B reaction above $E_p$ $\sim$ 25 MeV. This line shape component will be briefly discussed in the section giving results of the line shape calculations (section III).

Proton inelastic scattering to states of a collective band like the 4.439-MeV state of $^{12}$C generally has a strong direct reaction component. In such cases, calculations in the coupled-channels framework are often able to describe satisfactorily total and differential cross sections (see e.g. \cite{Satchler} and references therein). This is readily done with nuclear reaction codes like Ecis \cite{Ecis}, that provide the necessary flexibility in the reaction parameter input to define the ground and excited states and their couplings, potentials, etc. A broad outline of the nuclear reaction calculations has already been given in \cite{lshape}, but some practical aspects will be given anyhow further below.

At proton energies below about 15 MeV, the CN component is clearly present, evidenced by the distinct peaks in the cross section of $^{12}$C(p,p$\gamma_{4.438}$)$^{12}$C (see Fig. \ref{Excf44}). The typical separation of the peaks suggests that they may be formed from isolated CN resonances with definite spin and parity and in fact, a good part of them can be identified with known  states in the compound nucleus $^{13}$N. It is therefore tempting to describe the inelastic scattering reaction in the region up to about $E_p$ = 15 MeV by a superposition of the direct reaction mechanism and isolated CN resonances. In the following, the calculations of the line shapes and $\gamma$-ray angular distributions for the CN and direct component will be more detailed.

\subsection{Compound-nucleus resonances}

The formalism for inelastic scattering reactions proceeding through an intermediate CN resonance, or CN state, has been taken from the monograph of Ferguson \cite{Ferguson}. Taking a CN state with spin $b$, its decay by particle emission with angular momentum $L$ to an excited state of the target nucleus with spin $c$, followed by a deexcitation $\gamma$ ray with multipolarity $L_{\gamma}$ to target state $d$, is described as a cascade with angular momenta:

\begin{equation} (1) ~~~  \vec{b} = \vec{c} + \vec{L}  ~~~~~ (2)~~~  \vec{c} = \vec{d} + \vec{L_{\gamma}}
\label{bcL}
\end{equation}

The most general formulation of the angular correlation is  given by equation (2.96) in Ferguson, implying a summation over 20 indices. In the present case, the deexcitation of the $c$ = 2$^+$, 4.439-MeV state to the $^{12}$C ground state with $d$ = 0$^+$ fixes $L_{\gamma}$  = $c$ = 2,  which reduces considerably the formula. For the line shape calculations, in the case of an unobserved deexcited recoil nucleus, equation (2.96) leads to: 

\begin{eqnarray}
W(\theta_p,\theta_{\gamma},\phi_{\gamma}) = \sum_{}^{} t_{kq}(b) ~ \epsilon^{\star}_{k_{L} q_{L}}(L L') ~ \epsilon^{\star}_{k_L{_{\gamma}} q_L{_{\gamma}}}(L_{\gamma} L'_{\gamma}) ~ f_c ~ M(cLb) ~ M(cL'b)^{\star} 
\label{W}
\end{eqnarray}

where the tensor $t_{kq}(b)$ describes the CN state, the $\epsilon^{\star}_{kq}$ are efficiency tensors for particle and $\gamma$-ray counters, $f_c$ is a factor containing the angular couplings, and $M$ are reduced matrix elements. Summation is over $k,q,k_L,q_L,k_{L_{\gamma}},q_{L_{\gamma}},L,L'$. The different terms are detailed in the appendix of the present paper, the general expression of $t_{kq}(b)$ is given in equation (2.98) in Ferguson.

The tensor  $t_{kq}(b)$ is the result of the reaction in the ingoing channel, the formation of the CN state with angular momenta:

\begin{equation}  \vec{s} + \vec{l_i}  = \vec{b}
\label{ab}
\end{equation}

Here, $s$ = $\frac{1}{2}^+$ is the proton spin (spin of $^{12}$C$_{g.s.}$ = 0$^+$), which results in a unique orbital angular momentum $l_i$  for a given $b$ of the CN state. Assuming an unpolarized proton beam along the laboratory system z axis,  only $q$ = 0 elements of  $t_{kq}(b)$ are nonzero, and  $t_{kq}(b)$  can be reduced to:

\begin{equation}
t_{k0}(b) = (-1)^{a-b+k} ~ \frac{\hat{b}^2 \hat{l_i}^2}{\sqrt{4\pi} } ~ <l_i 0, l_i 0 \mid k 0 > ~ W(l_i b l_i b ; a k) 
\label{tkq}
\end{equation}

where  $\hat{j}$ = $\sqrt{2j + 1}$, $<l_i 0, l_i 0 \mid k 0 >$ is a Clebsch-Gordan coefficient and $W(l_i b l_i b ; a k) $ a Racah coefficent.  $k$ runs from 0 to 2$l_i$, and $k$ = even, selected by the Clebsch-Gordan coefficient. 

Equation \ref{W} is thus completely defined by the angular momenta and their couplings, with the exception of the relative values of the reduced matrix elements $M(cLb) ~ M(cL'b)$. For the outgoing proton, $L$ is the result of spin-orbit coupling $\vec{L}$ = $\vec{l}$ + $\vec{s}$. Since  it can be supposed that $M(cLb)$ does not depend on the proton spin ($M(cLb) = M(clb)$), the probability for decay of the CN state by proton emission with orbital angular momentum $l$ is given by $W(l) \propto M(clb)M(clb)^{\star}$.  In the present studies, 2 different values of $l$ were sufficient to describe the CN reactions at all energies. Let $l_0$ be the smallest possible angular momentum in the decay channel,  the branching ratio 

\begin{equation} 
W_{l_0} := \frac{W(l_0)}{W(l_0) + W(l_0 + 2)} 
\label{W02}
\end{equation} 
is the only free parameter. For the line shape calculations, equation \ref{W} was calculated as a function of $\theta_p$, separately for $l_0$ and $l_0 + 2$. The differential cross section ${d\sigma}/{d\Omega}(\theta_p)$ can be obtained from equation (2.96) in Ferguson, by assuming unobserved $\gamma$ ray and excited nucleus (see Appendix).  Finally, the line shapes and $\gamma$-ray angular distributions of the CN state were taken as an incoherent sum of both angular momenta with the respective weight factors $W_{l_0}$ and (1 - $W_{l_0}$).

\subsection{Direct reactions}

The formalism has been taken from the monograph of Satchler \cite{Satchler}. It is similar to the above described one for CN resonances, but with a different division of the treatment of the reaction sequence. The first two steps in the CN case, equations \ref{ab} and  \ref{bcL}(1), leading to the excited state of $^{12}$C by inelastic scattering, is here contracted to one step with spins 

\begin{equation} \vec{a} + \vec{s}  + \vec{l}_i \rightarrow \vec{b} + \vec{s} + \vec{l}_f \end{equation}

where $a$ and $b$ are the $^{12}$C ground and excited state, respectively, $s$ is the proton spin and $l_i$ and $l_f$ the incoming and outgoing angular momenta. 

The orientation of $b$ is the result of coherent summing of all contributions with possible combinations of $l_i$, $l_f$ and  $s$. It  is calculated in the framework of nuclear reactions in the coupled-channels formalism. Practically, the program Ecis97 \cite{Ecis} was used for these calculations. 

The angular distribution of the emitted $\gamma$ ray  is given by equation (10.132) of Satchler:

\begin{equation}
W(\theta_{\gamma},\phi_{\gamma}) ~ = ~ \sum_{kq}^{} t_{kq}(b,\theta_p) ~ R_k(\gamma) ~ \frac{\sqrt{4\pi}}{\hat{k}} ~ Y_{kq}(\theta_{\gamma},\phi_{\gamma}) 
\label{Wdir}
\end{equation}

where the polarization tensors $t_{kq}(b,\theta_p)$  describe the state of orientation of the excited state $b$ after proton inelastic scattering with angle $\theta_p$, and $R_k(\gamma)$ are the gamma-radiation parameters, detailed by equation (10.154) in Satchler (see Appendix).

The polarization tensors $t_{kq}(b,\theta_p)$ can be constructed from the scattering (or transition) amplitudes  $T$ (equation (10.32b) in Satchler):

\begin{equation}
t_{kq}(b,\theta_p) ~ = ~ \sum ~ T_{\beta,\sigma_o,\alpha,\sigma_i} ~ T^{\star}_{\beta+q,\sigma_o,\alpha,\sigma_i} ~ (-1)^{b-\beta} ~ \hat{b} ~ <b (\beta+q) b (-\beta) \mid k q> ~ (d\sigma/d\Omega(\theta_p))^{-1}
\label{TAmp}
\end{equation}

where $\alpha$, $\beta$  are the projections of spins $a$  and $b$, respectively and $\sigma_{i,o}$ the projections of the proton spin in the ingoing and outgoing channels. Summation in the present case ($\alpha = a = 0$) is only over $\beta,\sigma_i,\sigma_o$. 

Scattering amplitudes are either obtained directly from the nuclear reaction codes, or have to be assembled from other output parameters. The latter is the case for the code Ecis97 \cite{Ecis}, where the  amplitudes can be derived from  the scattering-matrix elements (see Appendix.)

\section{Line shape calculations}

\begin{figure} \includegraphics[width=15 cm]{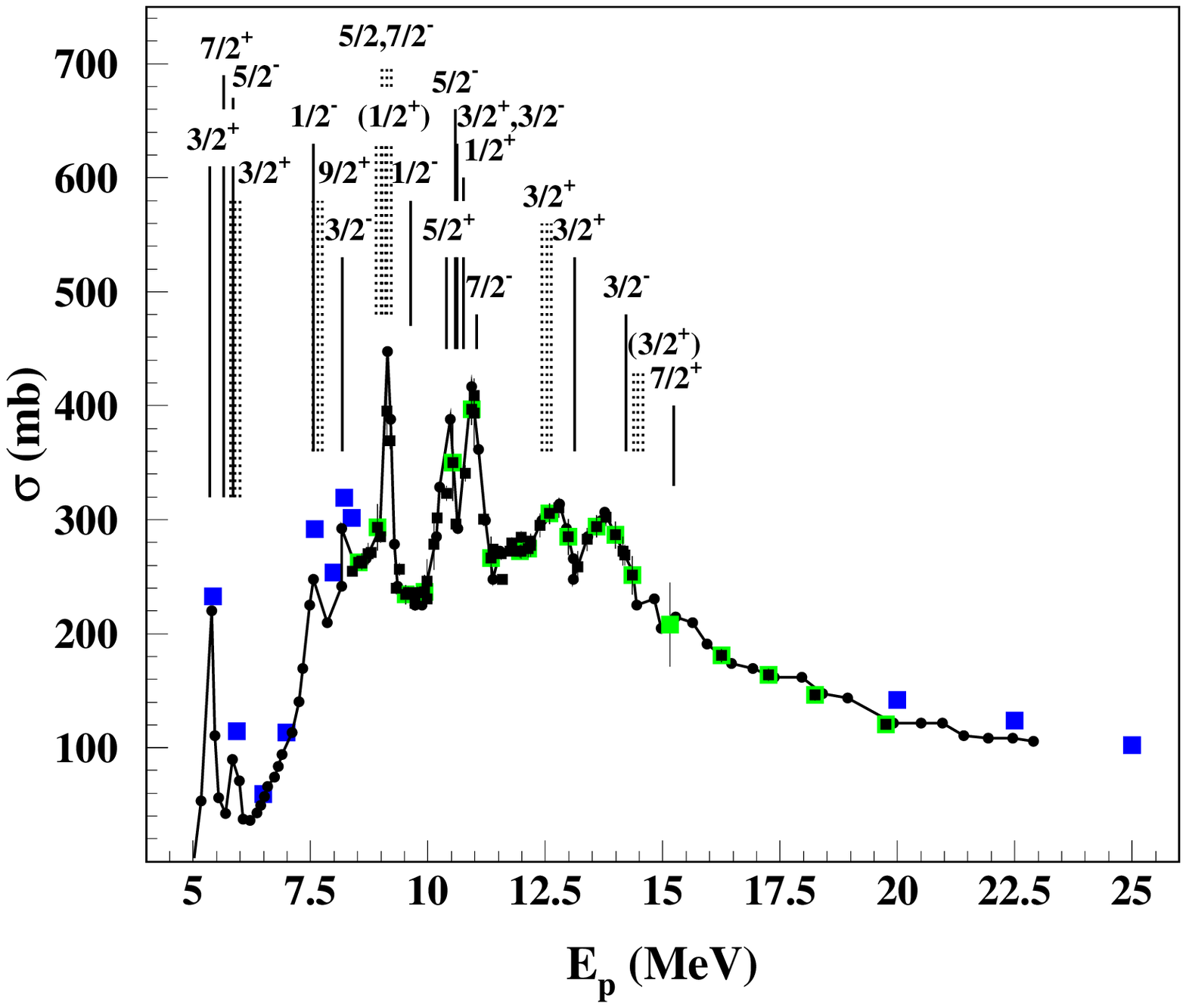}
\caption{(Color online) Cross section as a function of laboratory proton energy for emission of the 4.438-MeV $\gamma$ ray in proton inelastic scattering off $^{12}$C. Black dots connected with the black line are the data of Dyer et al. \cite{Dyer81}, small black squares the data of the Orsay-1997 experiment \cite{torion}. Larger squares indicate  Orsay data where line shapes are available: green for the 1997- and blue for the 2002-experiment \cite{t2002}. The data point at 15.15 MeV does not figure in  ref. \cite{torion}. Its large error bar reflects the uncertainty of the contribution of reactions with $^{16}$O present in the collodion target. Vertical lines indicate the position of  $^{13}$N states in the range $E_x$ = 6.5 - 18 MeV with known spin and parity. Broad hatched lines indicate states with either large uncertainties on the excitation energy ($>$100 keV or uncertainty not quoted) or very large widths ($>$1 MeV) (values taken from NuDat at NNDC ref. \cite{NNDC}).  } 
\label{Excf44}
\end{figure}

Measured line shapes come from essentially two experiments, both done at the tandem-Van-de-Graaff accelerator at Orsay. Line shapes from the first experiment in 1997 \cite{torion}  have already been used for the line shape studies of Kiener, de S\'er\'eville and Tatischeff \cite{lshape}. Line shape data are available for selected runs in the range $E_p$ = 8.4 - 15.2 MeV with collodion targets (chemical composition C$_{12}$H$_{16}$N$_4$O$_{18}$) and with carbon targets in the range  $E_p$ = 16.25 - 19.75 MeV. The energy ranges below $E_p$ = 8.4 MeV and above $E_p$ = 20 MeV were covered in the second experiment in 2002 \cite{t2002} with carbon targets.  Angular distribution data are available at some selected energies from the experiments of Dyer et al. \cite{Dyer81PC}. An overview of availabe line shape data and measured cross sections  is shown together with known resonance states in $^{13}$N in Fig. \ref{Excf44}. Additionally to that, some line shapes are published in Kolata, Auble \& Galonsky \cite{Kolata67} at $E_p$ = 23 MeV and Lang et al. \cite{Lang87} at $E_p$ = 40 MeV. 

In both Orsay experiments, $\gamma$-ray spectra have been obtained with large-volume coaxial HP-Ge detectors equipped with active BGO shielding. In the 1997 experiment \cite{torion}, 8 detectors were used in the angular range $\theta$ = 45$^{\circ}$ - 145$^{\circ}$, and 5 detectors in the 2002 experiment \cite{t2002} in the range $\theta$ = 45$^{\circ}$ - 157.5$^{\circ}$ in a first phase and after a detector failure, 4  detectors in the range $\theta$ = 67.5$^{\circ}$ - 157.5$^{\circ}$ in a second phase. The HP-Ge crystals had typically 7 cm diameter, 7 cm length and were at about 35 cm from the target.   Except for the two highest proton beam energies $E_p$ = 22.5 and 25 MeV, high-statistics profiles of the 4.439-MeV line on a relatively small background have been recorded.  For the comparison with calculated line shapes, the Compton background below the lines was subtracted by linear interpolation (1997 experiment) or with the help of Geant simulations  of the detection setup (2002 experiment).

Line shape calculations for CN resonances and direct reactions were realized with two separate Monte-Carlo type programs, very similar except for the formalism describing the angular correlations. The programs consisted in event-by-event simulation of the reaction sequence  from the inelastic scattering reaction to the emission and detection of the $\gamma$ ray. The interaction depth in the target, center-of-mass (CM) scattering angle $\theta_p$, time of the $\gamma$-ray emission and $\gamma$-ray emission angles $\theta_{\gamma},\phi_{\gamma}$, were randomly drawn from cumulative distributions reflecting the different probabilities. For the $\gamma$-ray emission, the probabilities $W(\theta_{\gamma},\phi_{\gamma})$ depend on the CM scattering angle. Therefore, a series of cumulative ($\theta_{\gamma},\phi_{\gamma}$)-distributions  were made for a grid of CM scattering angles with $\Delta \theta$ = 5$^{\circ}$ in the case of CN resonances and $\Delta \theta$ = 10$^{\circ}$ for direct reactions.

Recoil angle and energy of the excited $^{12}$C in the laboratory were determined from nuclear reaction kinematics following the CM scattering angle and incoming proton energy. The slowing down of the recoil $^{12}$C  in the target was calculated with the stopping power tables of SRIM \cite{SRIM} until the deexcitation of the 4.439-MeV state and emission of the $\gamma$ ray with a half-life of $t_{1/2}$ = 61 fs occurred. Angular straggling was ignored as well as the geometrical effect of the finite beam-spot size on the target. Polar angle and energy of the emitted $\gamma$ ray in the laboratory were calculated with relativistic kinematics and stored in an energy-angle histogram.  Line shapes were then constructed from the energy-angle histogram for  the solid angles spanned by the HP-Ge detectors employed in the Orsay experiments. Only full-energy events in the detectors were considered, with the effect of detector energy resolution on the line shapes, typically 4.5 keV FWHM at 4.4 MeV, included. A few million events per proton energy were simulated, resulting in high-statistics calculated line profiles.

For the direct reaction component, differential cross sections and angular correlations were obtained with the nuclear reaction code ECIS \cite{Ecis} in the coupled-channels formalism. Rotational coupling between the $^{12}$C 0$^+$ ground state and the 2$^+$ state at 4.439 MeV with deformation parameters $\beta_2$ = -0.61 and  $\beta_4$ = 0.05 was used. At energies above $E_p$ = 15 MeV, the 4$^+$ state at 14.08 MeV was also included. This coupling scheme was taken from Meigooni et al. \cite{Meigooni85}, who used it in coupled-channels calculations to derive an energy- and channel-dependent potential for nucleon scattering off $^{12}$C in the range $E_{n}$ = 20-100 MeV. Calculations with this potential were already succesfully used in the previous study of Kiener, de S\'er\'eville and Tatischeff \cite{lshape} for the direct reaction component above $E_p$ = 15 MeV. It provided also good results at lower energies, except at a few energies below $E_p$ = 10 MeV. There, other potentials from the compilation of Perey and Perey \cite{Perey}  were also tried and sometimes provided a better description of the data. 

Line shapes and $\gamma$-ray angular distributions of the CN component at each proton energy were calculated with the spins and parities $J^{\pi}$ of close by $^{13}$N states.  It provided mostly satisfactory descriptions of data in the region of dominating CN component with a few exceptions discussed below.  At some proton energies, where several nearby states exist, calculations with all of the $J^{\pi}$'s  were made and compared to the data. Calculations with the different spin parities showed in general marked differences in the line shapes or the $\gamma$-ray angular distributions, such that the choice was usually evident.

After selection of the CN state $J^{\pi}$, the branching ratio $W_{l_0}$ of CN state decay orbital angular momenta and the proportion of CN ($\equiv$$W_{CN}$) and direct reaction component (1-$W_{CN}$) were scanned, aiming for a simultaneous good reproduction of the line shapes in all detectors and the $\gamma$-ray angular distribution. Practically, for each $W_{l_0}$, varied in steps in steps of 0.05 - 0.2, the parameter $W_{CN}$ was found in a fit of the angular distribution and the resulting line shapes calculated and compared to the data.  Usually, the calculated line shapes at most or all detector angles and the angular distribution had their best agreement with experiment for very similar parameter values. 

\section{Results}

\subsection{Orsay-1997 experiment}

The results of the parameter search for the CN and the used nuclear potential for the  direct component to reproduce the measured line shapes and $\gamma$-ray angular distributions are summarized in table \ref{tab97}. 
Good agreement with measured data could be obtained at each energy with the potential of Meigooni et al. \cite{Meigooni85} for the direct reaction component and one  single resonance $J^{\pi}$ for the CN component. At all energies except the lowest, the $\chi^2$(C) obtained with the calculated angular distributions are similar to and sometimes even smaller than the $\chi^2$(L) of Legendre-polynomial fits to the angular distribution data. The $\chi^2$ are in both cases obtained with 3 free parameters ($W_{l_0}$,  $W_{CN}$ and an overall normalization factor for the calculated distributions, 3 coefficients for the Legendre-polynomial fit of an electric quadrupole transition 2$^+$ $\rightarrow$ 0$^+$). Error bars on the experimental data are of the order of 4 - 6 \%.
 
The $J^{\pi}$ of the best adjustment corresponds in nearly all cases to a resonance state in $^{13}$N whose energy is within $\Gamma_{tot}$, the total width of the state, to the excitation energy $E_x$ in the CN reaction. At $E_p$ = 12.0 MeV ($E_x$ = 13.0 MeV), calculations with $J^{\pi}$ = $\frac{3}{2}^+$  corresponding to the most probable state at 13.50(20) MeV ($\Gamma_{tot}$ $\sim$6500 keV) could not reproduce the data. They were reproduced by a $J^{\pi}$ = $\frac{7}{2}^-$, that may be attributed to the state at 12.937(24) MeV ($\Gamma_{tot}$ $>$ 400 keV) whose $J^{\pi}$ is not known.  No nearby $\frac{5}{2}^+$  state in $^{13}$N was found at $E_p$ = 16.25 MeV, but the measured data could also be reasonably reproduced without a CN component. 

As expected, the proportion of the CN component $W_{CN}$ is high for proton energies on the three narrow peaks at 9.1, 10.4 and 10.9 MeV in the cross section excitation function, see Fig. \ref{Excf44}. For the peak at 9.1 MeV, the closest measurement is at $E_p$ = 9.2 MeV, where, however, no line shapes were available. Because of the particular importance of the cross section here, the result of the angular distribution fit was nonetheless included in table \ref{tab97}. There, equivalent $\chi^2$ were obtained for $\frac{1}{2}^+$ and $\frac{7}{2}^-$ resonances, but $\frac{1}{2}^+$ is more probable because the width of  the $^{13}$N state with $J^{\pi}$ = $\frac{7}{2}^-$ ($E_x$ = 10.360 MeV, $\Gamma_{tot}$ = 75 keV) seems barely compatible with the apparent peak width of $\sim$200 keV in the cross section data of Dyer et al. \cite{Dyer81}.  

Below $E_p$ = 12 MeV with significant CN component, the values of $W_{l_0}$ and $W_{CN}$ are typically constrained to within $\pm$0.15. For $\frac{1}{2}^-$ resonances, decay angular momenta $l$ = 1 and $l$ = 3 give exactly the same line shapes and $\gamma$-ray angular distributions, such that $W_{l_0}$ could not be constrained. For  $\frac{1}{2}^+$ resonances, only $l$ = 2 is possible. Above that energy, the CN component is weak, reasonable adjustements could also be obtained throughout with $W_{CN}$ = 0, and the constraints on $W_{l_0}$ are weaker ($\pm$ $\sim$0.3). At the highest energy with the collodion target, $E_p$ = 15.2 MeV, the $^{16}$O($\alpha$,2$\alpha$)$^{12}$C reaction contributes and has been added in the calculations.

\begin{figure} \includegraphics[width= 8 cm]{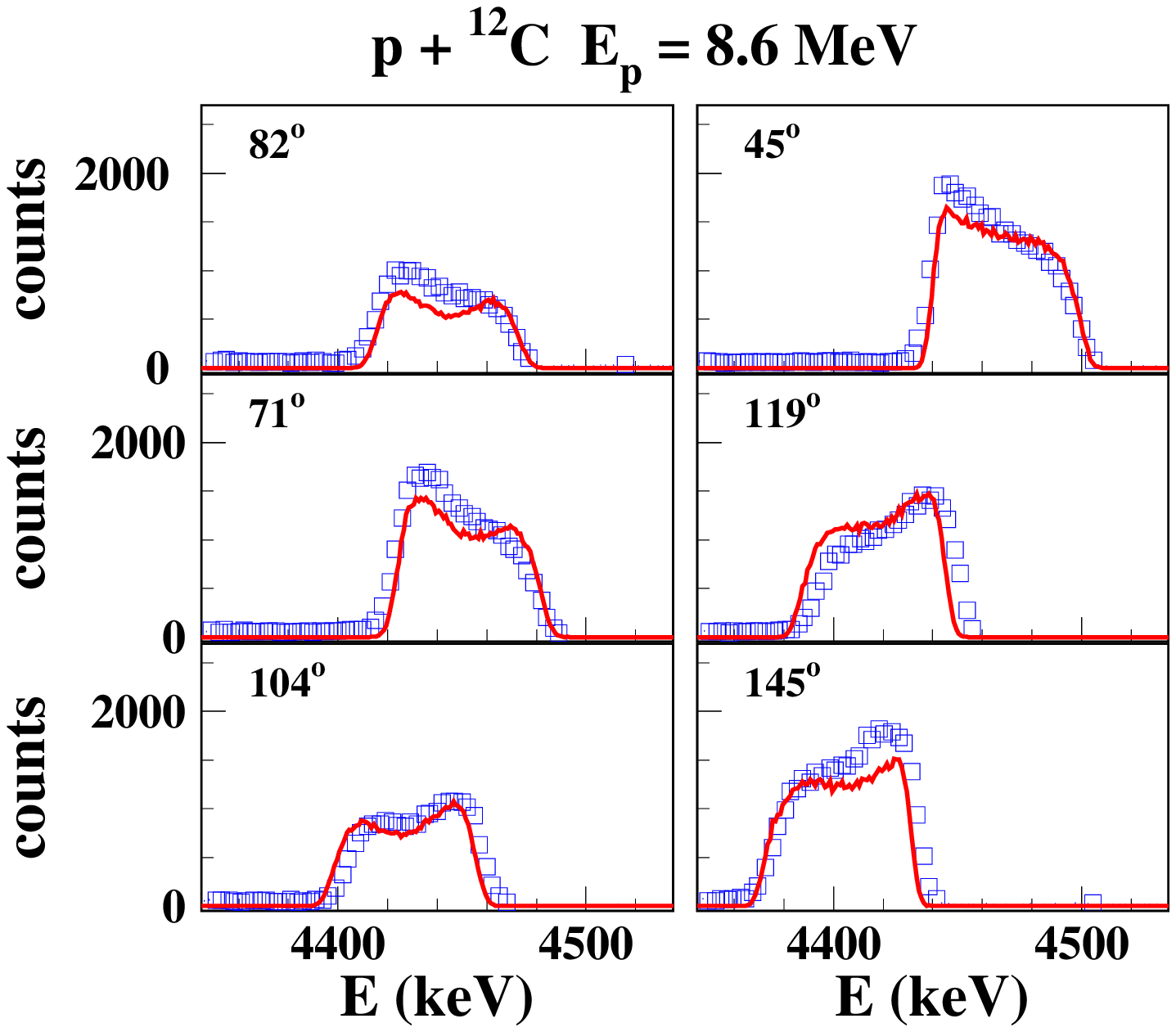} \includegraphics[width=8 cm]{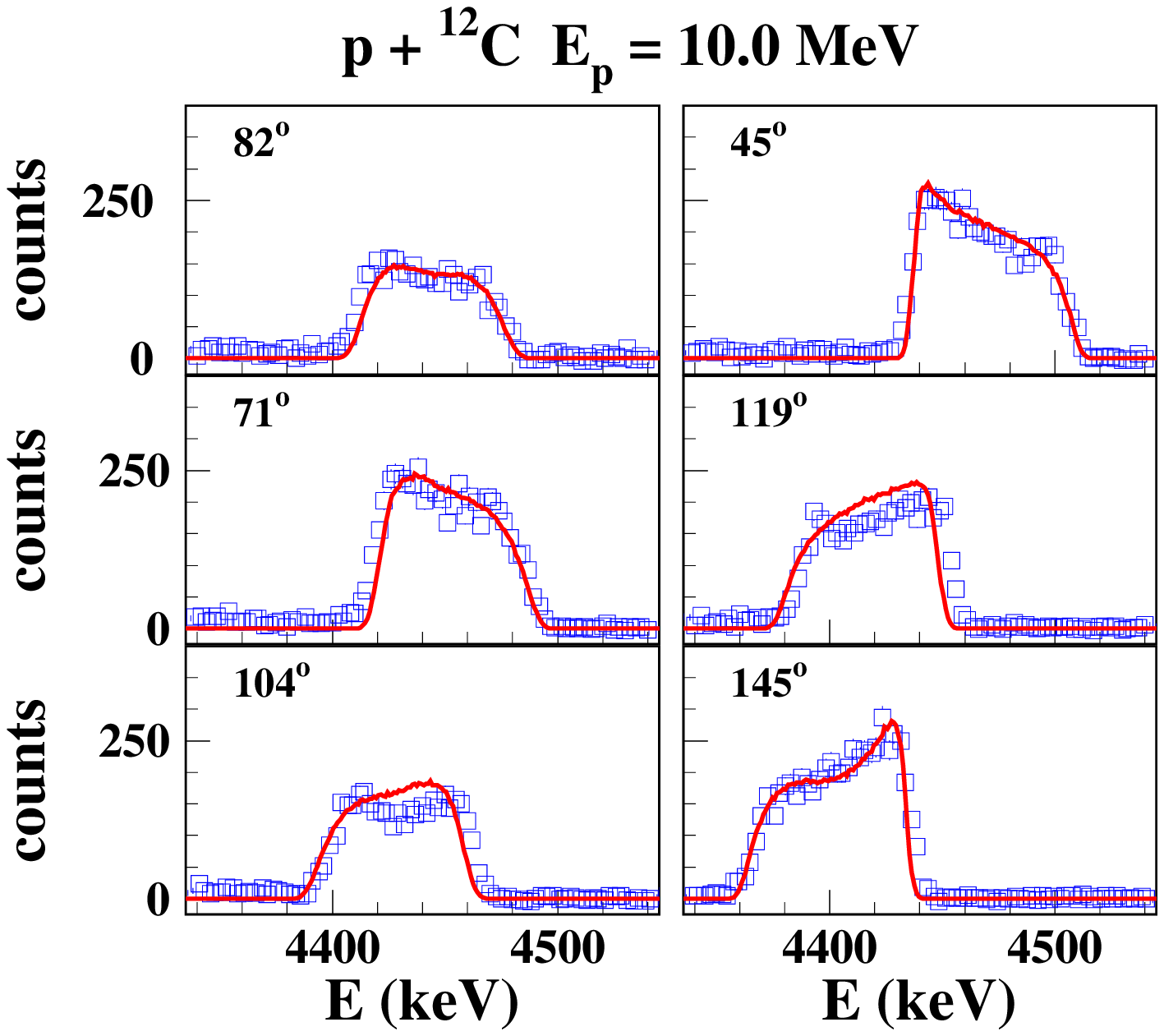}  \includegraphics[width=8 cm]{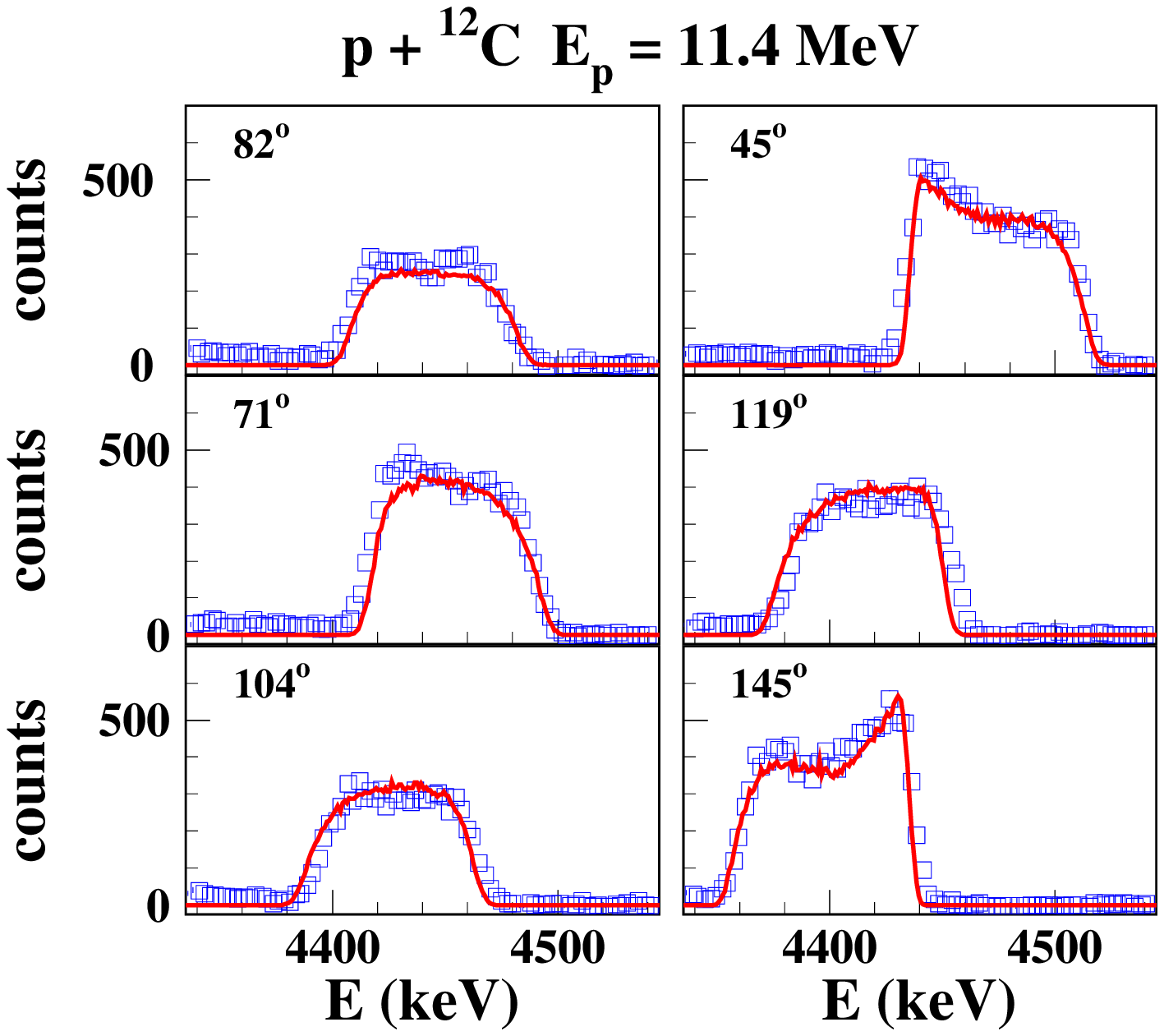}  \includegraphics[width=8 cm]{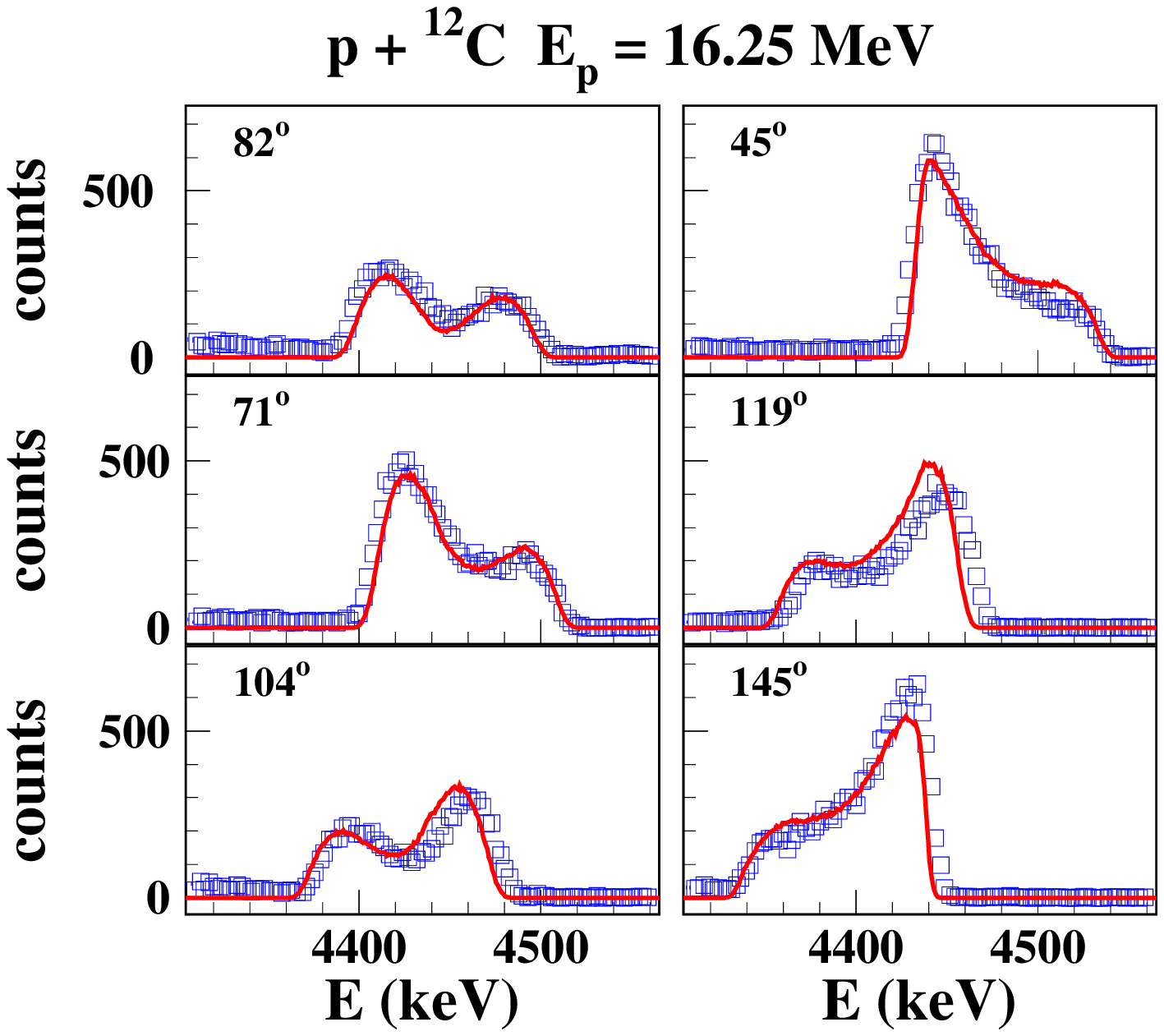} 
\caption{(Color online) Measured shapes of the 4.439-MeV $\gamma$-ray line from proton-inelastic scattering off $^{12}$C in the Orsay-1997 experiment  (blue symbols) and results of the line shape  calculation with parameters of table \ref{tab97} (red lines) at the proton beam energies indicated on the figures. } 
\label{Shape97}
\end{figure}

\begin{figure} \includegraphics[width= 7.5 cm]{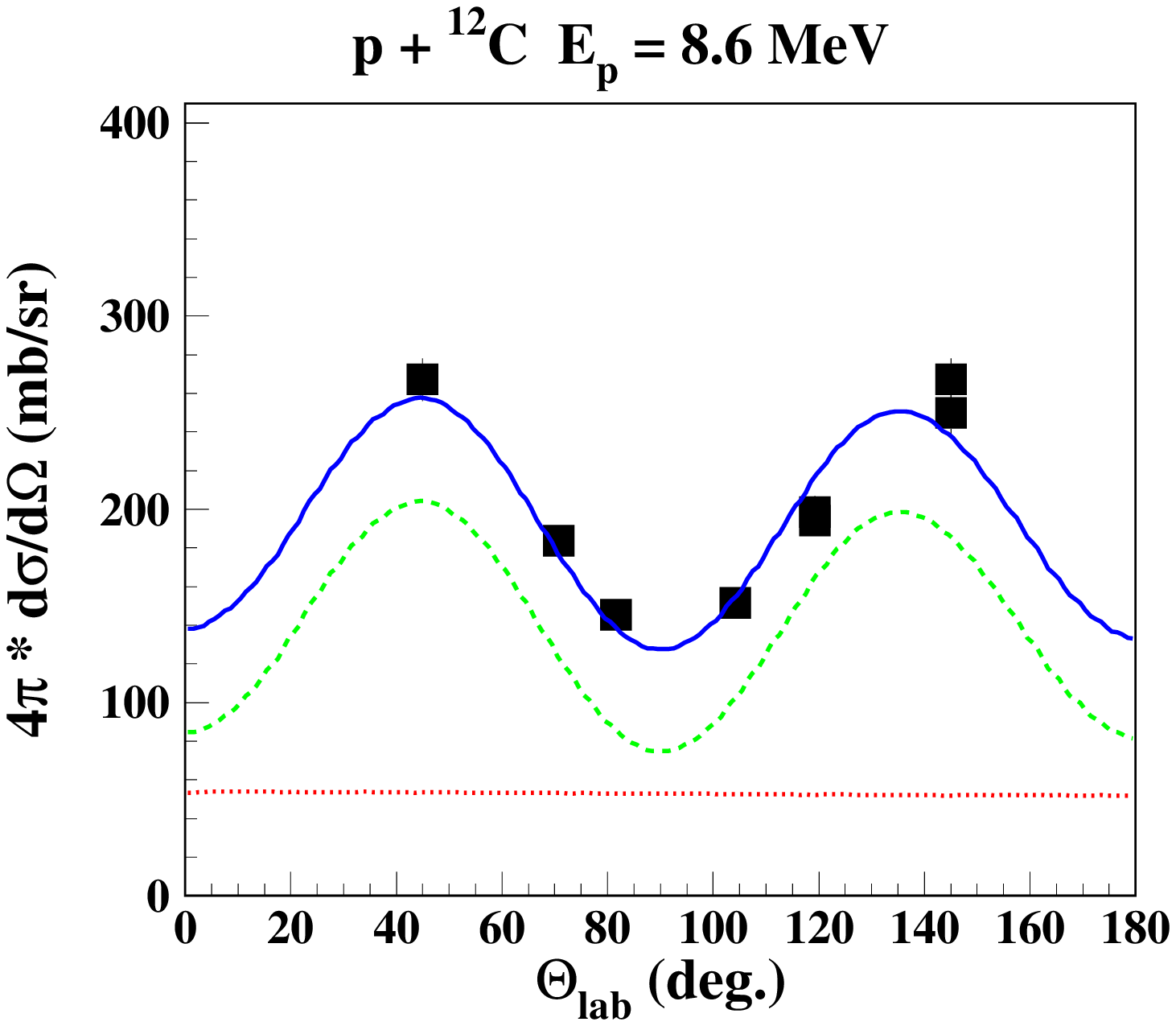} \includegraphics[width=7.5 cm]{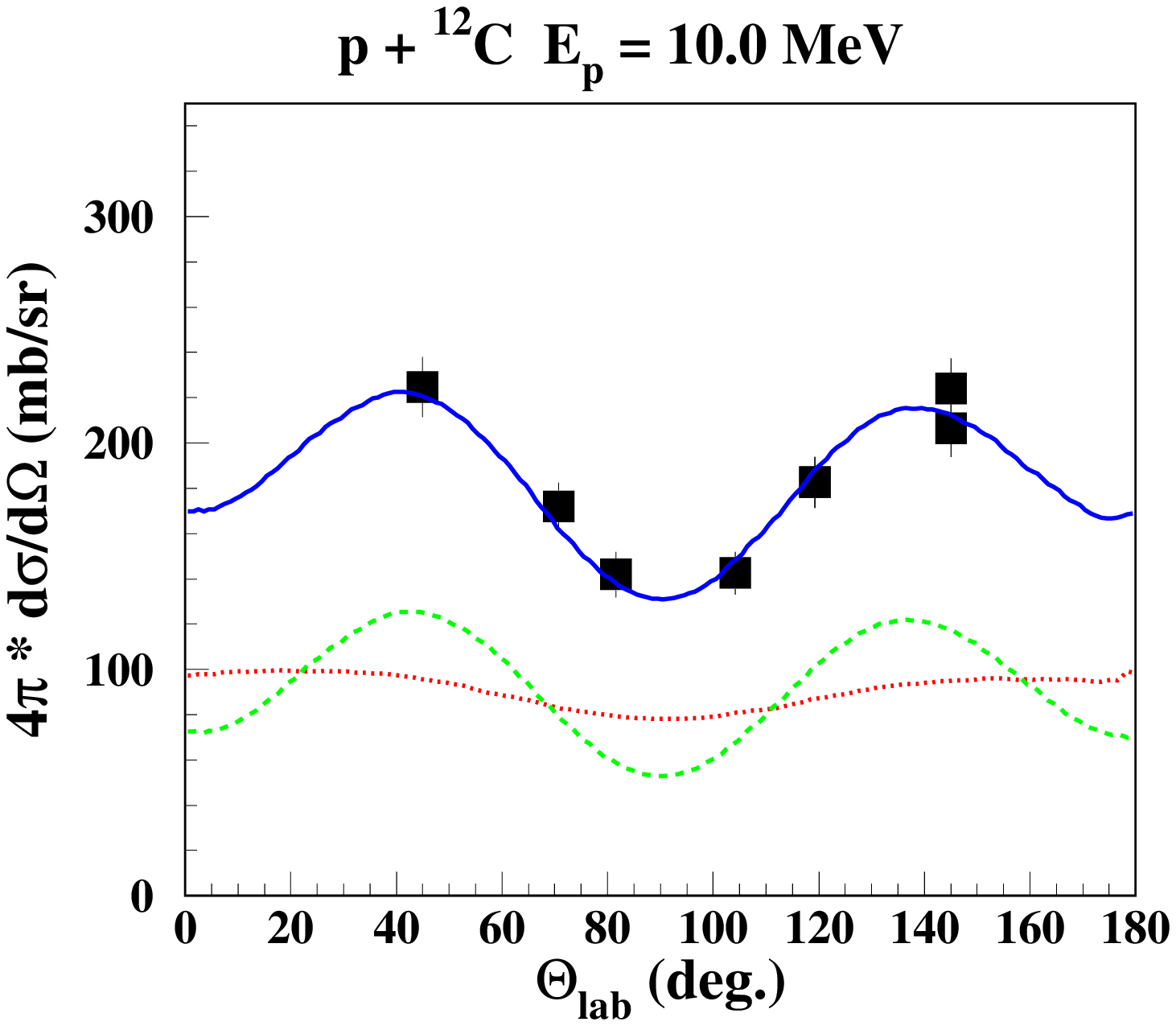}  \includegraphics[width=7.5 cm]{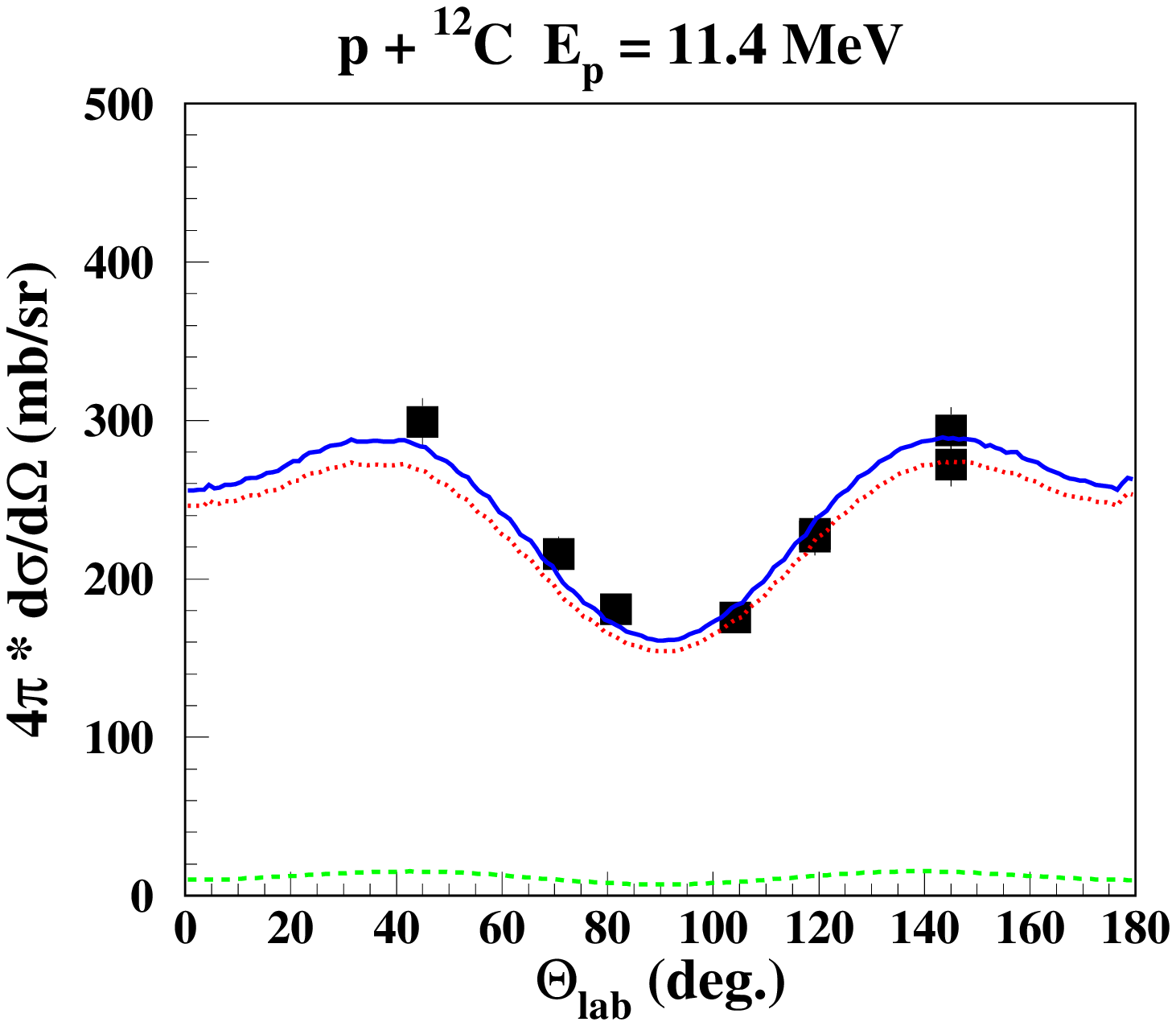}  \includegraphics[width=7.5 cm]{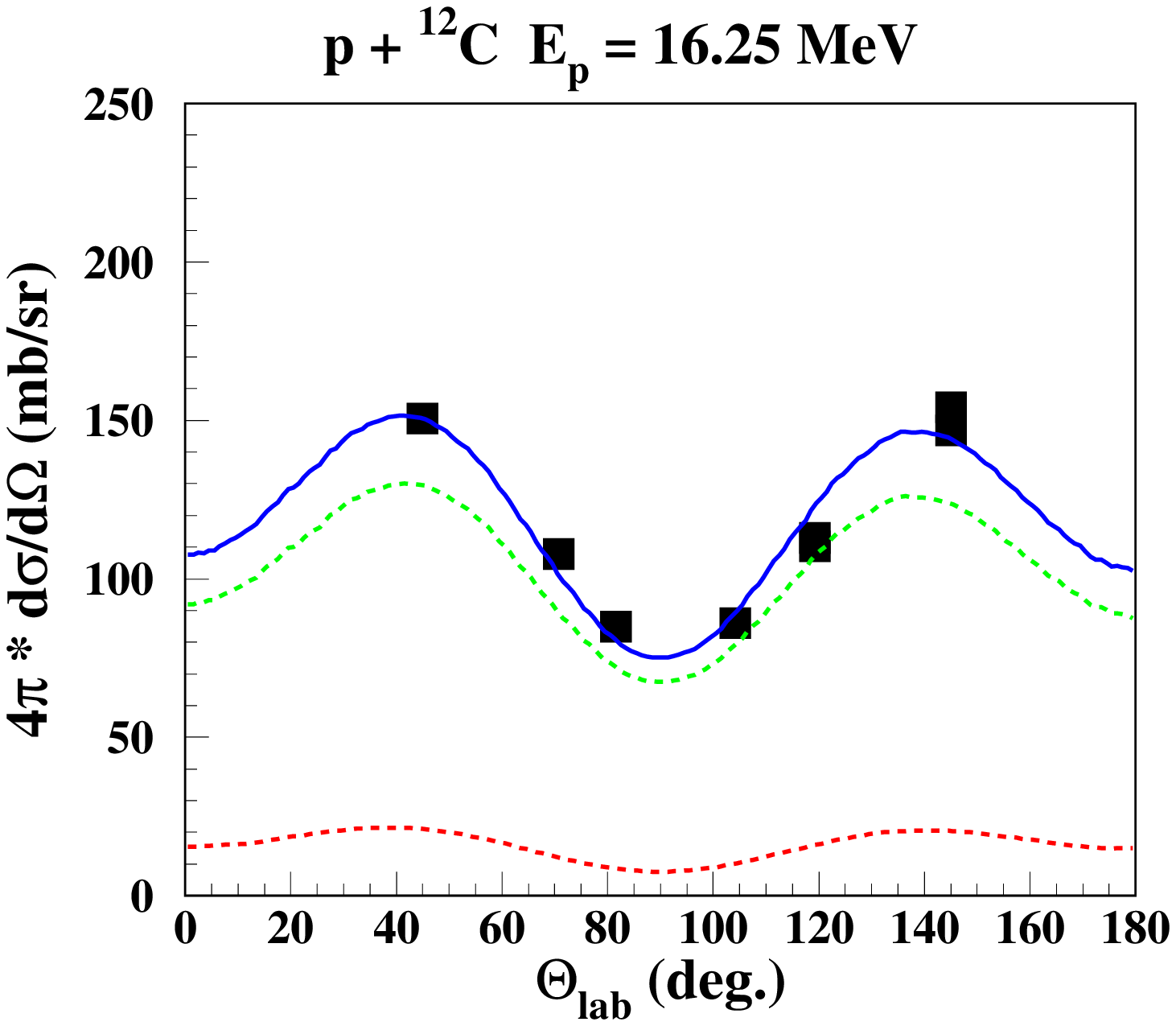} 
\caption{(Color online) Measured angular distribution data of the 4.439-MeV $\gamma$-ray line from proton-inelastic scattering off $^{12}$C in the Orsay-1997 experiment are presented by black symbols. The results of the calculations with parameters of table \ref{tab97} are shown by dotted red, dashed green and solid blue lines for the CN and direct component, and the sum of both, respectively. Proton beam energies are indicated on the figures.  } 
\label{Dist97}
\end{figure}

Calculated and measured line shapes and angular distributions at 4 proton beam energies are shown in Figs.  \ref{Shape97} and \ref{Dist97}. It includes proton energies with largely dominating CN or direct reaction components, at $E_p$ = 11.4 MeV and $E_p$ = 16.25 MeV, respectively, and energies with a more equal mixture of  both components at $ E_p$ = 8.6 MeV and at $E_p$ = 10.0 MeV. These examples represent also the typical degree of agreement between calculation and measured data, with very good adjustments of the $\gamma$-ray angular distribution and reasonably well reproduced line shapes with some deviations of the order of 20\% at a few angles.
 
\subsection{Orsay-2002 experiment}

The results of the parameter search to reproduce measured line shapes and $\gamma$-ray angular distributions in the  energy ranges $E_p$ = 5.44 - 8.4 MeV and $E_p$ = 20.0 - 25 MeV, are listed in table \ref{tab02}.   Good agreements with measured data could be obtained with the direct reaction and one single resonance $J^{\pi}$ component. In the range $E_p$ = 5.44 - 8.4 MeV, the angular distribution adjustments are good to excellent, considering error bars on the measured data of 1.5 - 2.5 \%. For the three highest energies, experimental error bars are of the order of 5 - 15 \% due to lower count statistics, and adjustments are good. 

The potential of Meigooni et al. \cite{Meigooni85} gave good results in the $E_p$ = 20-25 MeV range but was not able to reproduce several of the experimental data below 8 MeV. Attempts with modified strengths of the surface imaginary potentials and different coupling schemes could in some cases improve the agreement with data, but were not conclusive. Also attempts with a low-energy extension \cite{Wep17} of another global nucleon-nucleus optical model \cite{Wep09}, did not give a better description of the data. Finally, for the sake of simplicity, it was decided to use optical model potentials listed in the compilation of Perey and Perey \cite{Perey} for similar proton energies (see table \ref{tab02}), and without modification in the coupled-channels calculations. This could significantly improve the agreement with the experimental data for 4 proton energies.

For the CN component below $E_p$ = 7.6 MeV, the best agreements were obtained with $J^{\pi}$'s corresponding to the most probable $^{13}$N state. At $E_p$ = 8.0 - 8.4 MeV, calculations with $J^{\pi}$ = $\frac{9}{2}^+$ and $\frac{3}{2}^-$ of the closest  $^{13}$N states at 9.000 and 9.476(8) MeV, respectively, were unable to describe the data. They could, however, be reproduced with a $\frac{3}{2}^+$ component, probably corresponding to the  $\frac{3}{2}^+$ state at 7.900 MeV whose  total width is $\Gamma_{tot}$ $\sim$1500 keV. Above $E_p$ = 20 MeV, the data could best be described with a $\frac{5}{2}^-$ CN component. A corresponding state with known spin parity could only be found for $E_p$ = 20 MeV. Here, line shapes are better reproduced with a negligible CN component, while the angular distributions are better fitted with $\sim$30\% CN component ($\chi^2$(C) = 0.3-2.1). The indicated values $W_{CN}$ represent a compromise and are also in better agreement with adjustments of other angular distribution data in this energy range, see below.

Again, similar to the results for the Orsay-1997 data, there is a sizeable CN component at lower proton energies, while the data above $E_p$ = 20 MeV are  dominated by the direct reaction component. $W_{CN}$  and  $W_{l_0}$ are typically constrained within $\pm$0.15, except at  $E_p$ = 20 - 25 MeV, where the uncertainties on $W_{l_0}$ are of the order of $\pm$0.3.

\begin{figure} \includegraphics[width= 8 cm]{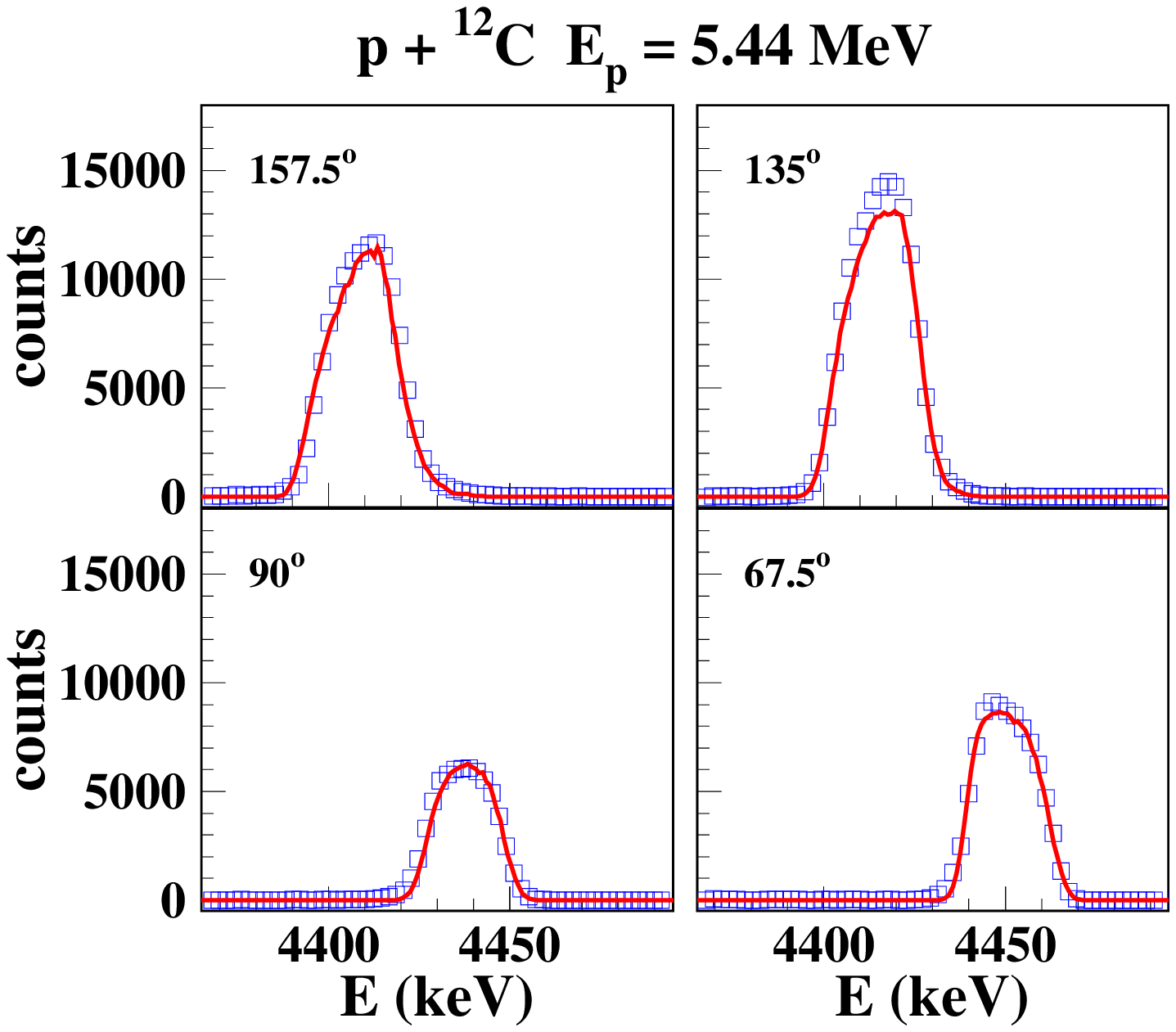} \includegraphics[width=8 cm]{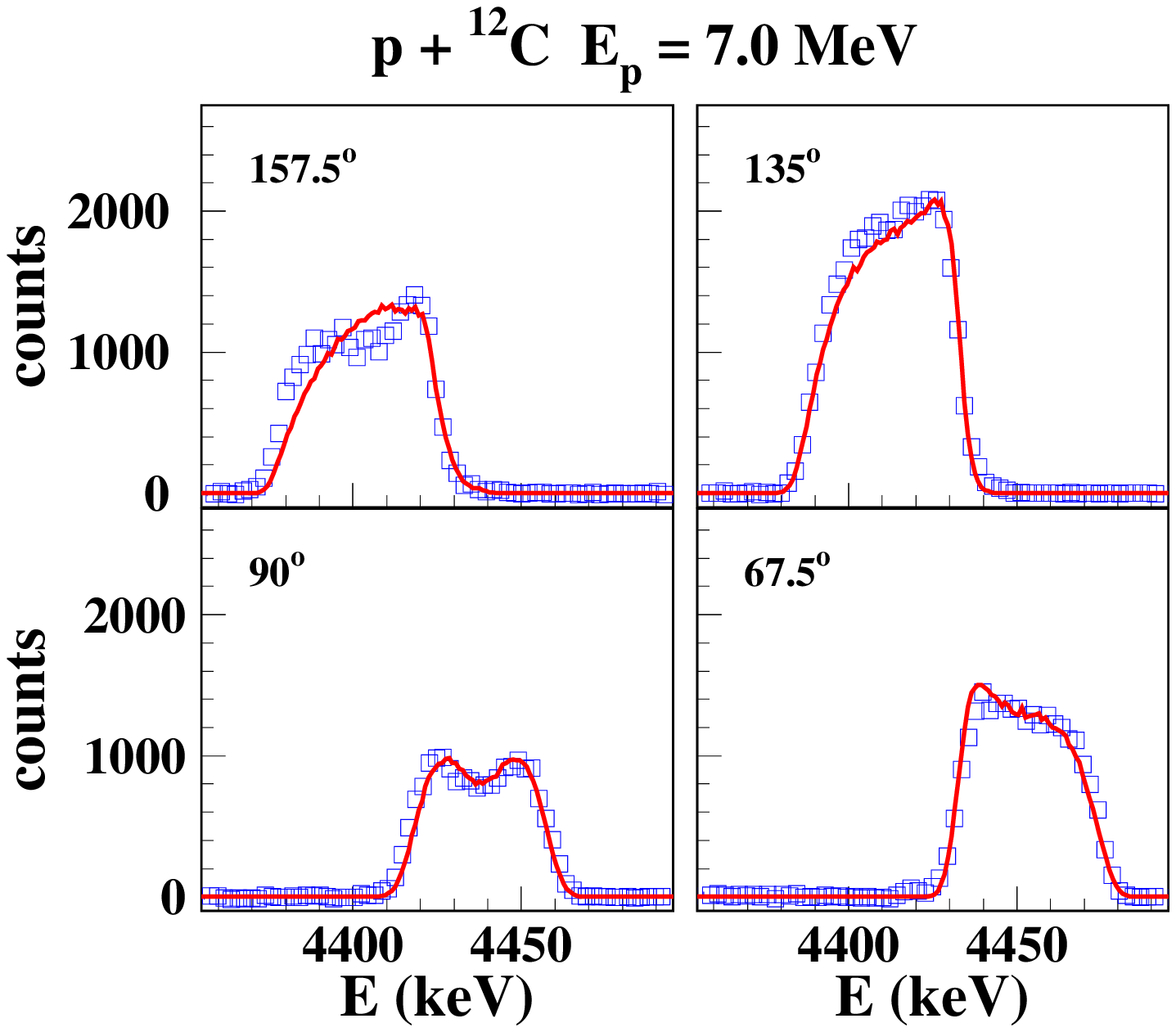}  \includegraphics[width=8 cm]{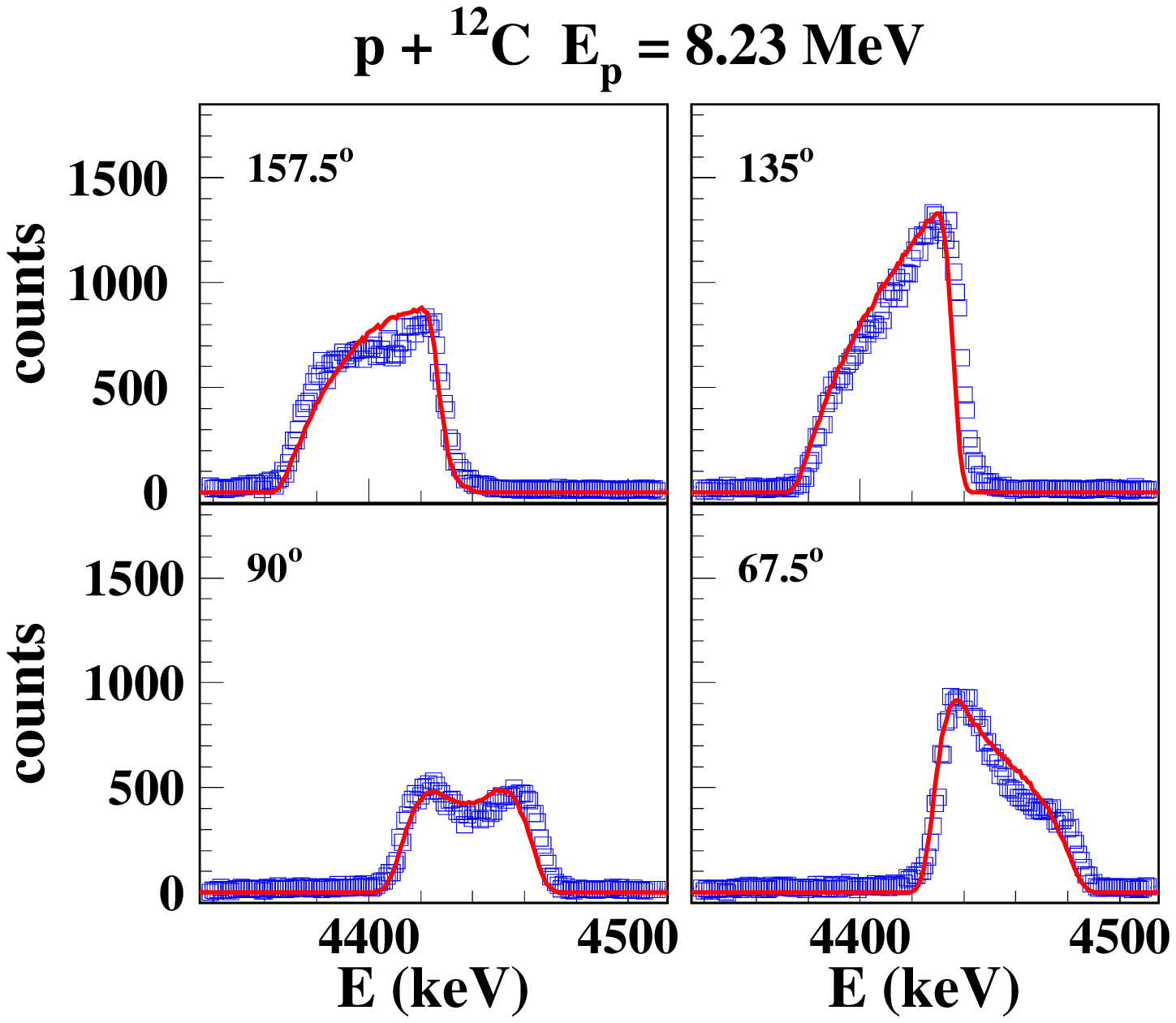}  \includegraphics[width=8 cm]{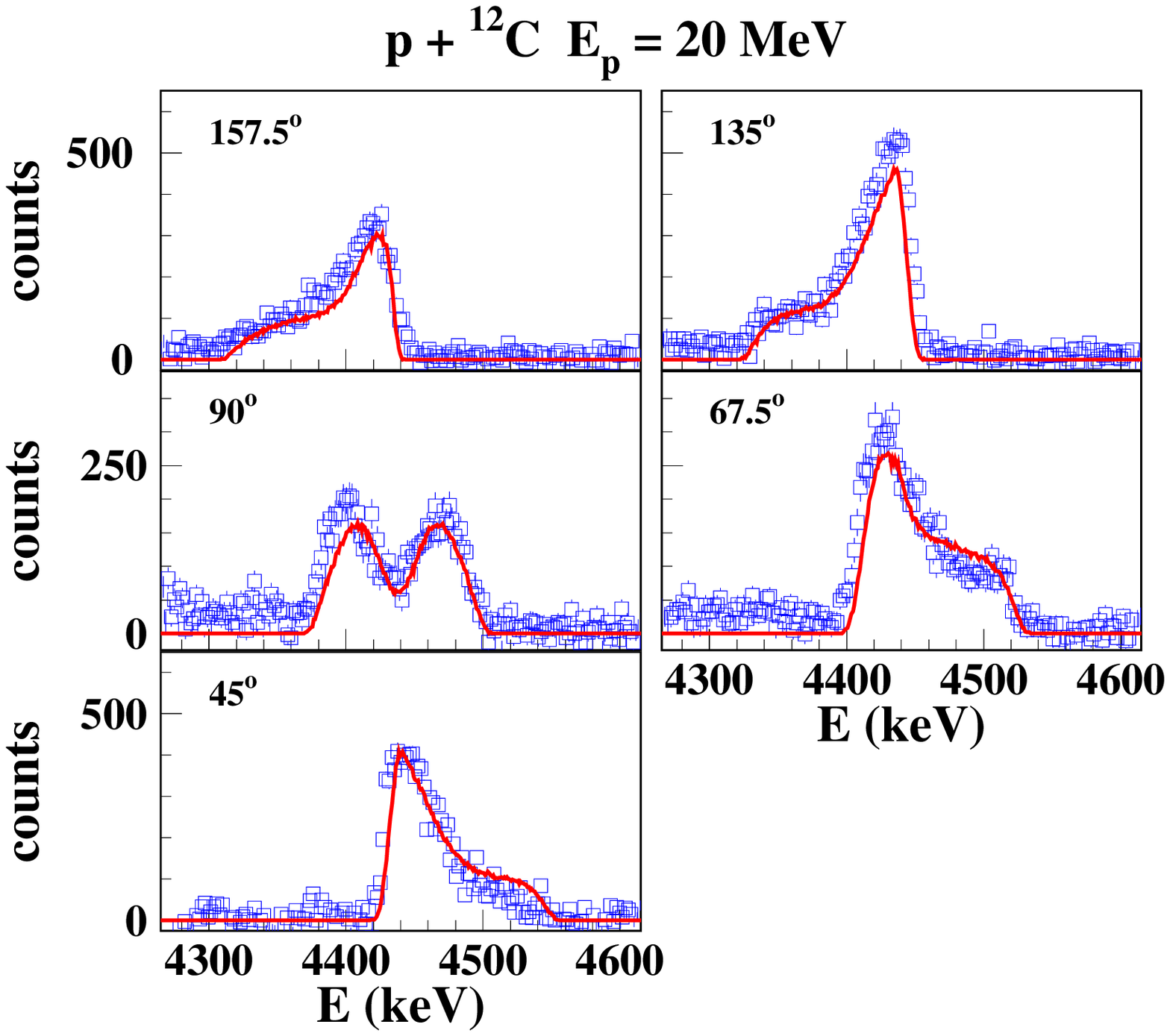} 
\caption{(Color online) Measured shapes of the 4.439-MeV $\gamma$-ray line from proton-inelastic scattering off $^{12}$C in the Orsay-2002 experiment  (blue symbols) and results of the line shape  calculation with parameters of table \ref{tab02} (red lines) at the proton beam energies indicated on the figures.  } 
\label{Shape02}
\end{figure}

\begin{figure} \includegraphics[width= 7.5 cm]{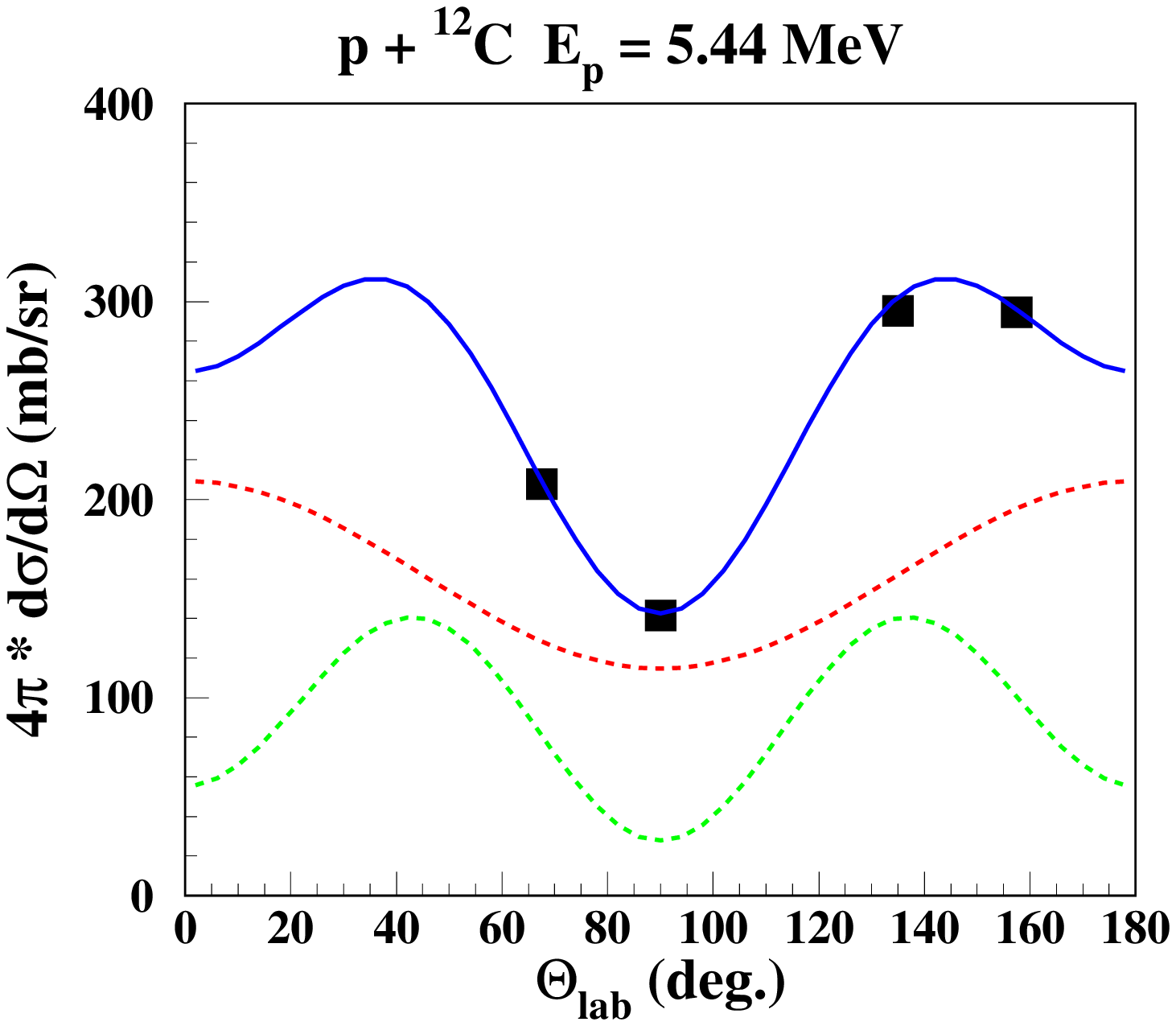} \includegraphics[width=7.5 cm]{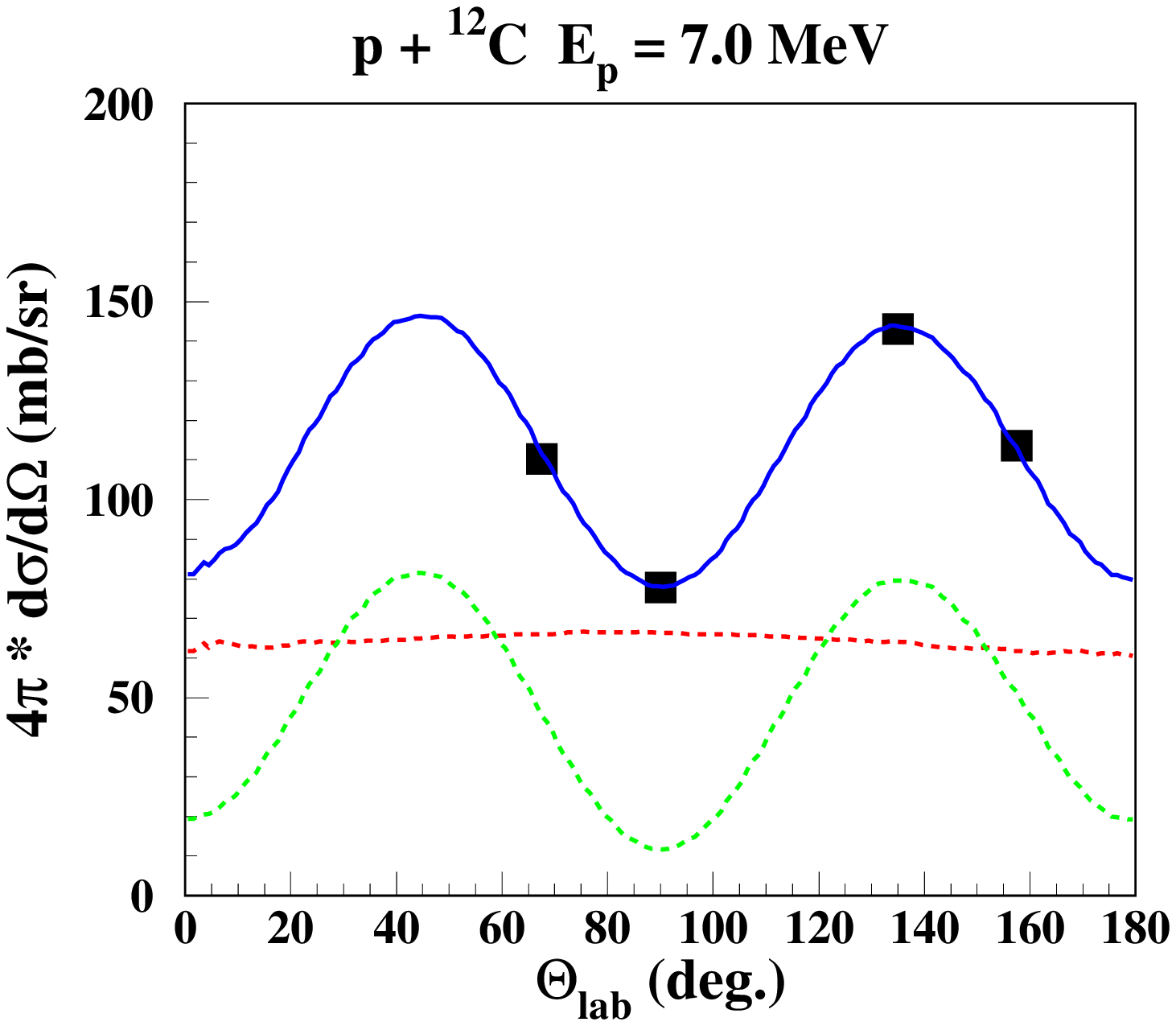}  \includegraphics[width=7.5 cm]{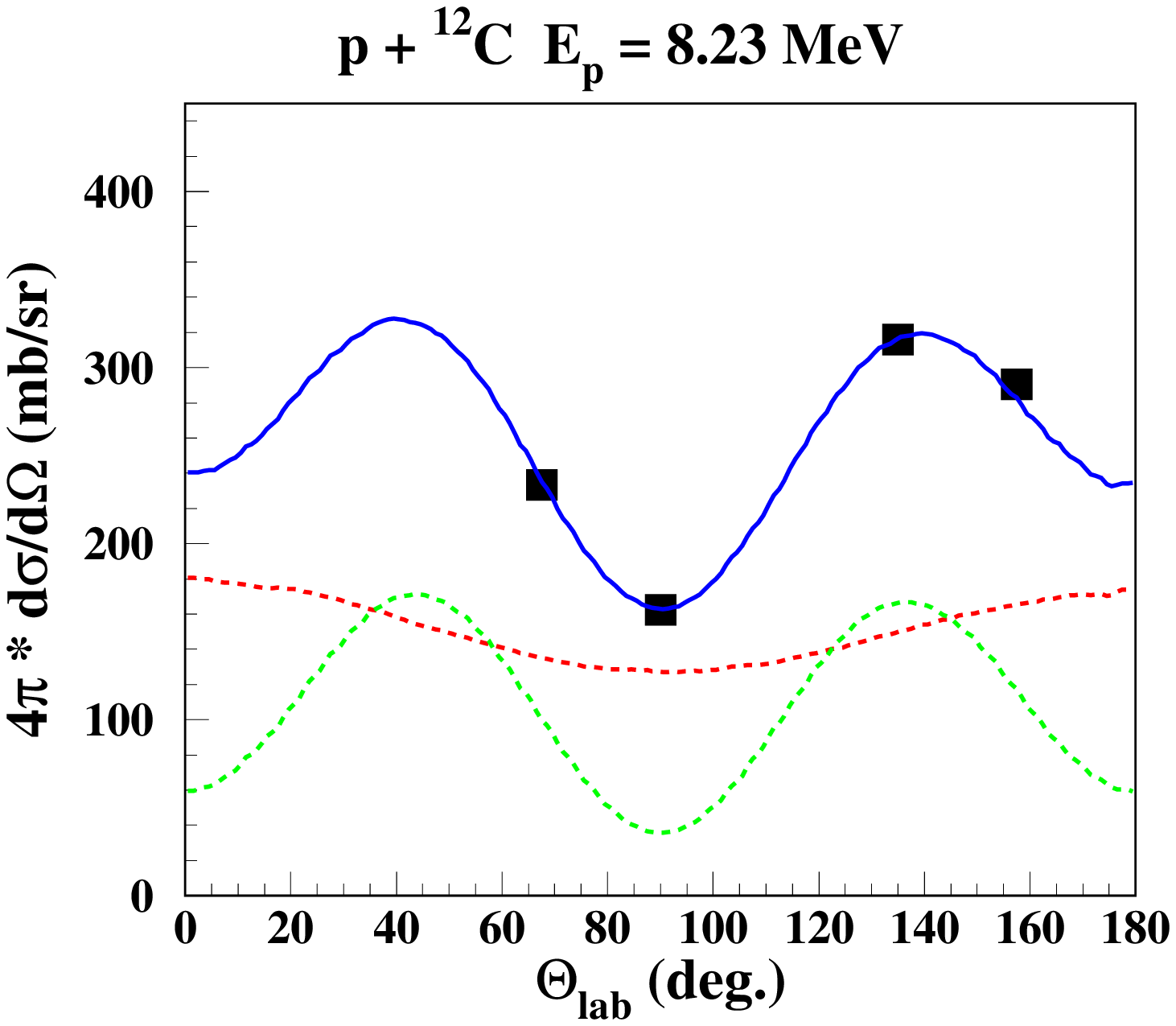}  \includegraphics[width=7.5 cm]{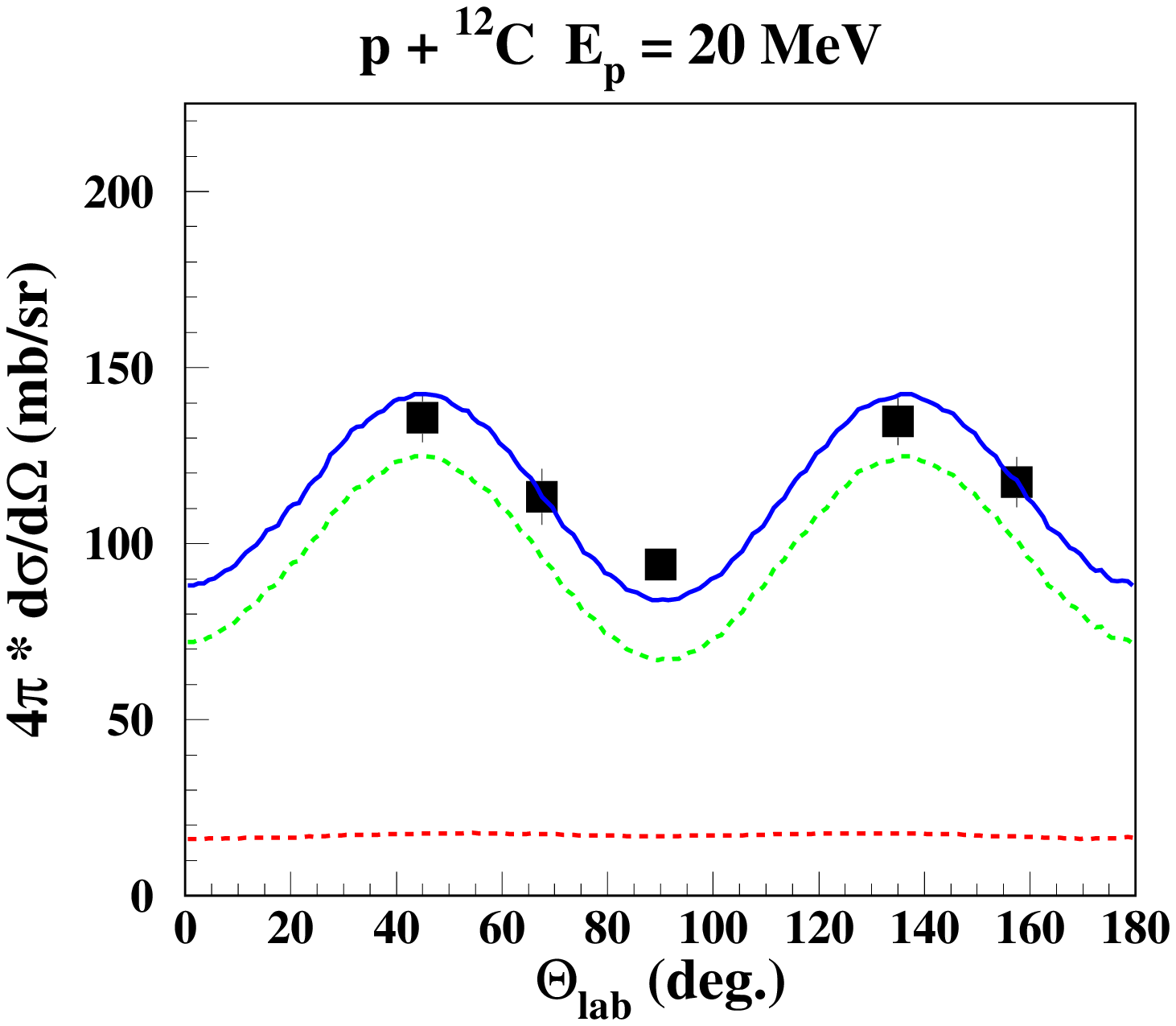} 
\caption{(Color online) Measured angular distribution data of the 4.439-MeV $\gamma$-ray line from proton-inelastic scattering off $^{12}$C in the Orsay-2002 experiment are presented by black symbols. The different lines have the meaning as in Fig. \ref{Dist97}. Proton beam energies are indicated on the figures.  } 
\label{Dist02}
\end{figure}

Calculated and measured line shapes and angular distributions at 4 proton beam energies are shown in Figs.  \ref{Shape02}, \ref{Dist02}. There is an excellent agreement between the calculated laboratory $\gamma$-ray angular distributions  and the measured data for the 3 lower proton energies. As mentioned above, the angular distribution at $E_p$ = 20 MeV would be better fitted with about twice as much of the quasi-isotropic CN component, however with a degrading fit for the line shapes. The calculated line shapes at $E_p$ = 5.44 MeV and 20 MeV reproduce nicely the experimental data, while for $E_p$ = 7.0 MeV and 8.23 MeV, the agreement is generally good, except at 157.5$^{\circ}$. 

\subsection{Other experiments}

\begin{figure} \includegraphics[width= 5.25 cm]{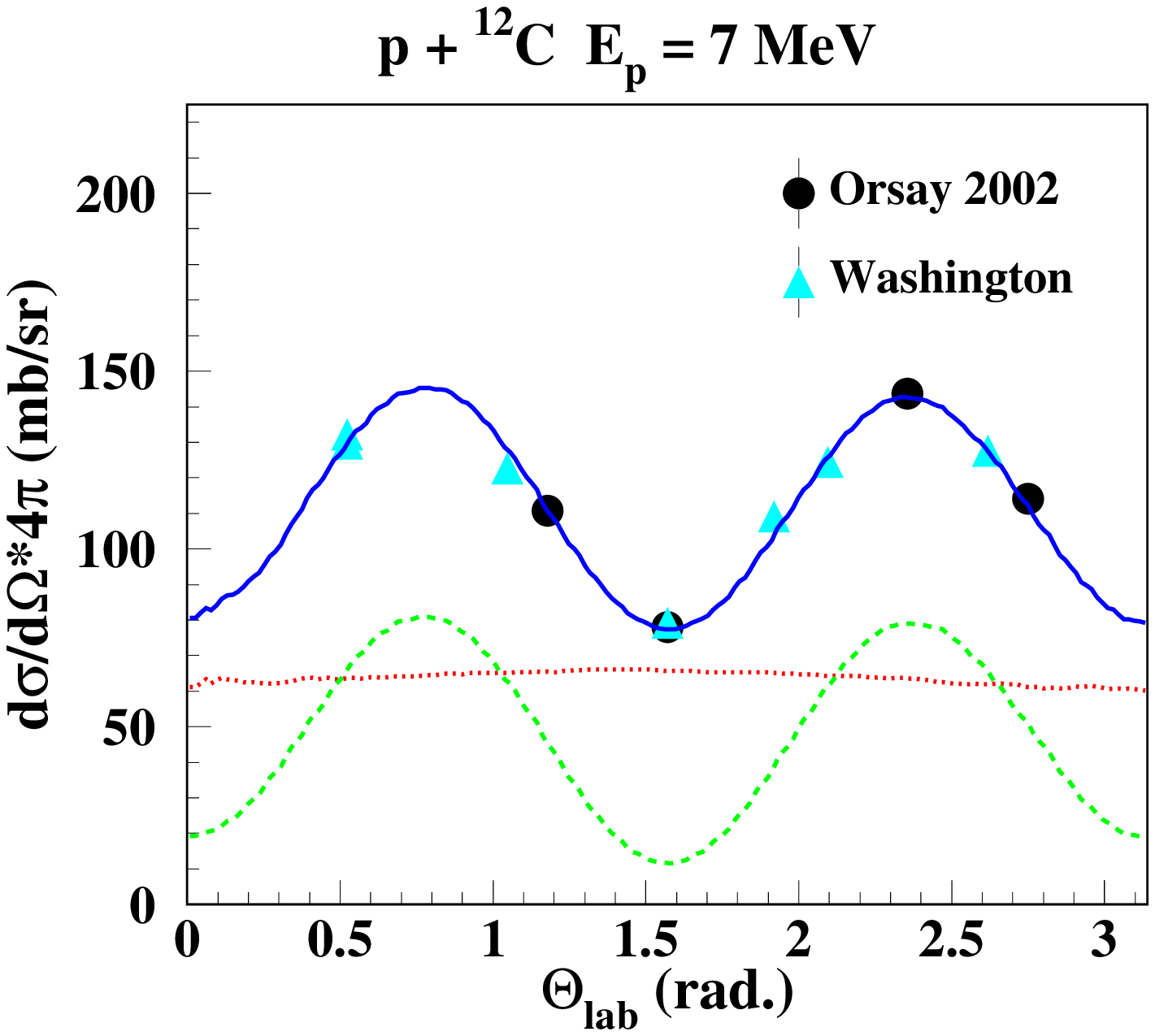} \includegraphics[width=5.25 cm]{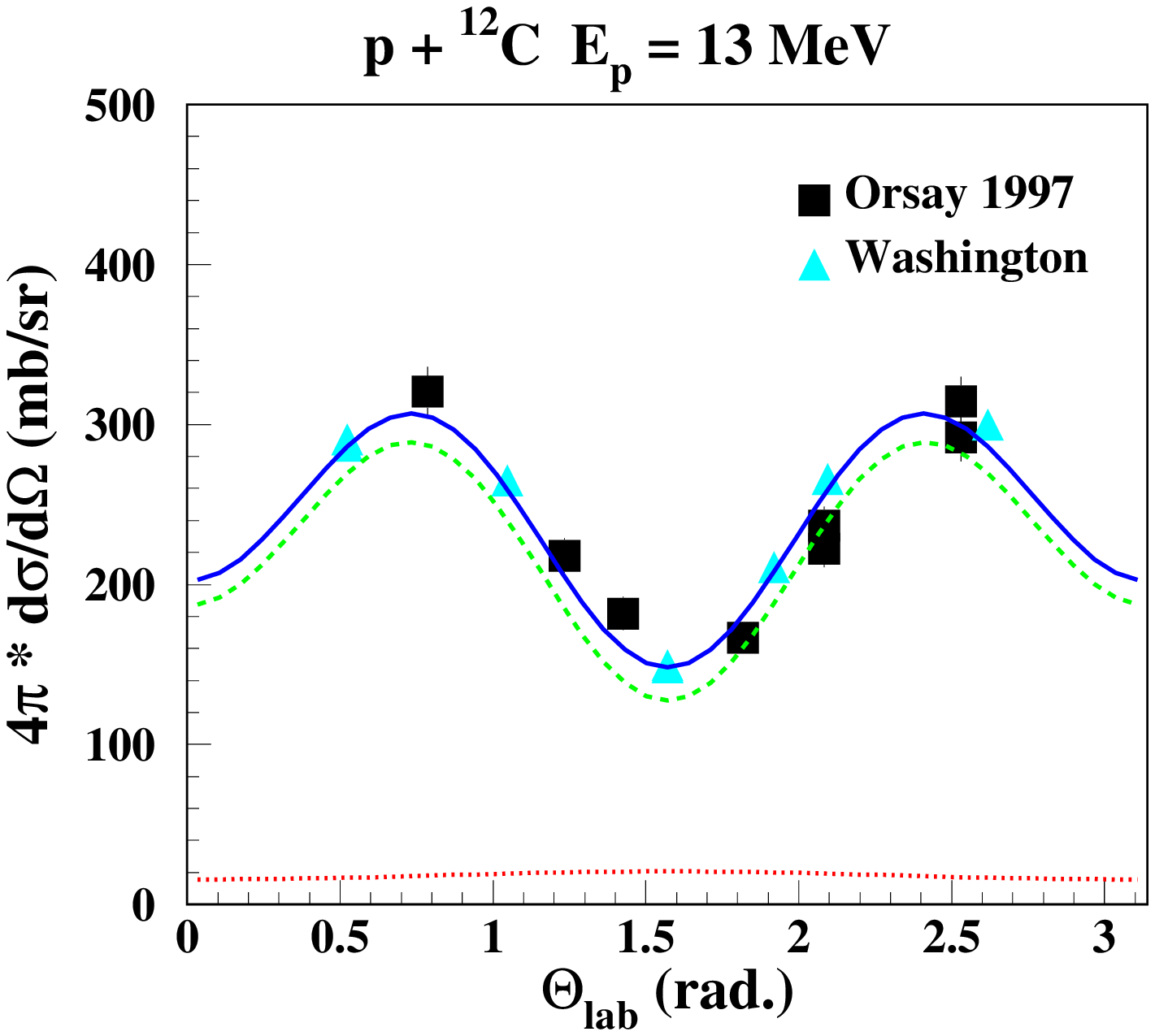}  \includegraphics[width=5.25 cm]{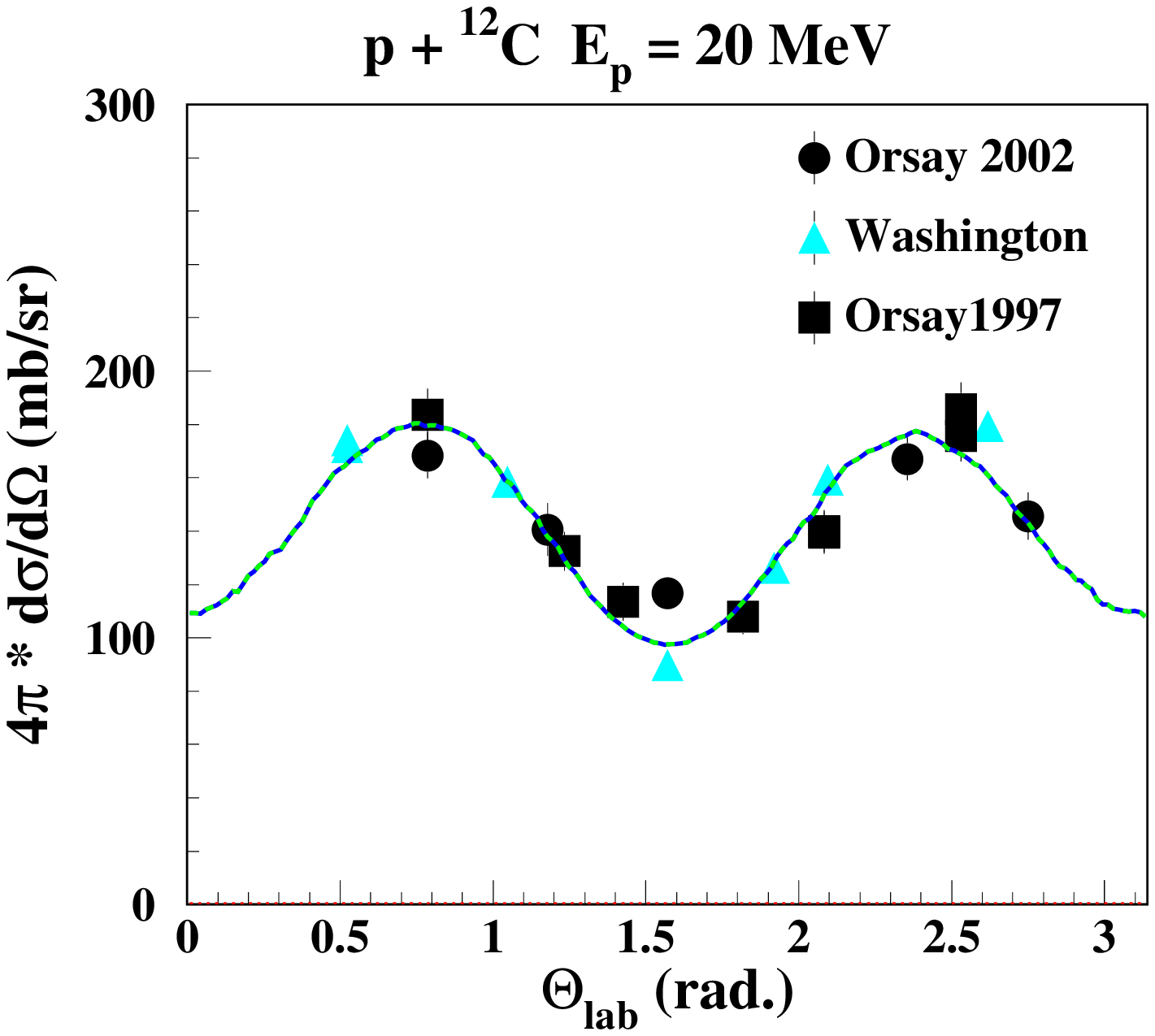}  
\caption{(Color online) Measured angular distribution data of the 4.439-MeV $\gamma$-ray line from proton-inelastic scattering off $^{12}$C from Orsay 1997 (black squares), Orsay 2002 (black circles)  and Washington (cyan triangles). The different lines have the same meaning as in Fig. \ref{Dist97}. Proton beam energies are indicated on the figures. For $E_p$ = 20 MeV, the data and calculated distributions at $E_p$ = 19.75 MeV from the Orsay-1997 experiment have been used.} 
\label{DistWO}
\end{figure}

The most comprehensive data set of 4.438-MeV $\gamma$-ray cross sections in proton inelastic scattering off $^{12}$C has been obtained at the Washington tandem accelerator from threshold up to $E_p$ = 23 MeV by Dyer et al. \cite{Dyer81}. No line shapes are available from that experiment, but angular distributions are available at some selected energies \cite{Dyer81PC}. They agree well with the Orsay-1997 and 2002 data up to $E_p$ = 18 MeV, and thus support the results presented in Tables \ref{tab97}, \ref{tab02}. At higher energies, the Washington and the Orsay-1997 data have slightly more pronounced minima and maxima than the Orsay-2002 data and clearly agree with negligible CN components. This is illustrated on  Fig. \ref{DistWO} where Orsay and Washington data scaled to the same integrated cross section, are shown together for some proton energies.

\begin{figure} \includegraphics[width= 7.5 cm]{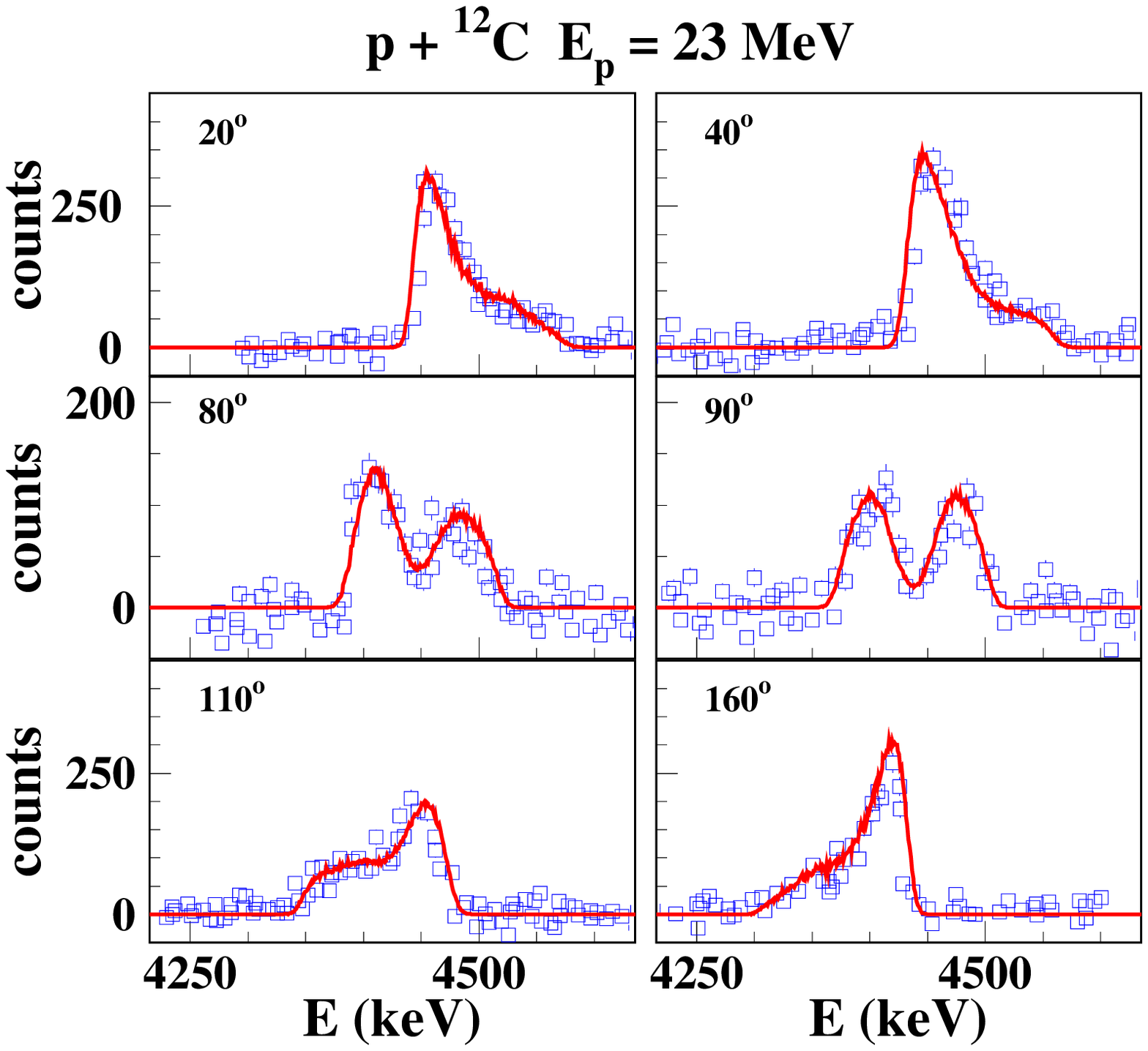} 
\caption{(Color online) Blue squares show measured shapes of the 4.439-MeV $\gamma$-ray line from proton-inelastic scattering off $^{12}$C at $E_p$ = 23 MeV  (data around 2$^{nd}$ escape peak from Kolata, Auble \& Galonsky \cite{Kolata67}, energy corrected and shifted by 1.022 MeV, see text) and results of the line shape calculation (red lines). } \label{ShapeK}
\end{figure}

Kolata, Auble \& Galonsky \cite{Kolata67} published partial $\gamma$-ray spectra around the second escape peak of the 4.438-MeV line at 11 different angles in the range $\theta_{lab}$ = 20 - 160$^{\circ}$  for $E_p$ = 23 MeV. These data, after subtracting  an estimated background and applying an energy shift of 40 keV for all detectors (in the spectra of Fig. of Ref. \cite{Kolata67}, only channel numbers are shown, and the energy dispersion is indicated in the caption), are well reproduced by the direct reaction component with the potential of Meigooni et al. \cite{Meigooni85} and with a negligible CN component. A comparison with calculated line shapes at 6 different angles is shown on Fig. \ref{ShapeK}.

\begin{figure} \includegraphics[width=7.5 cm]{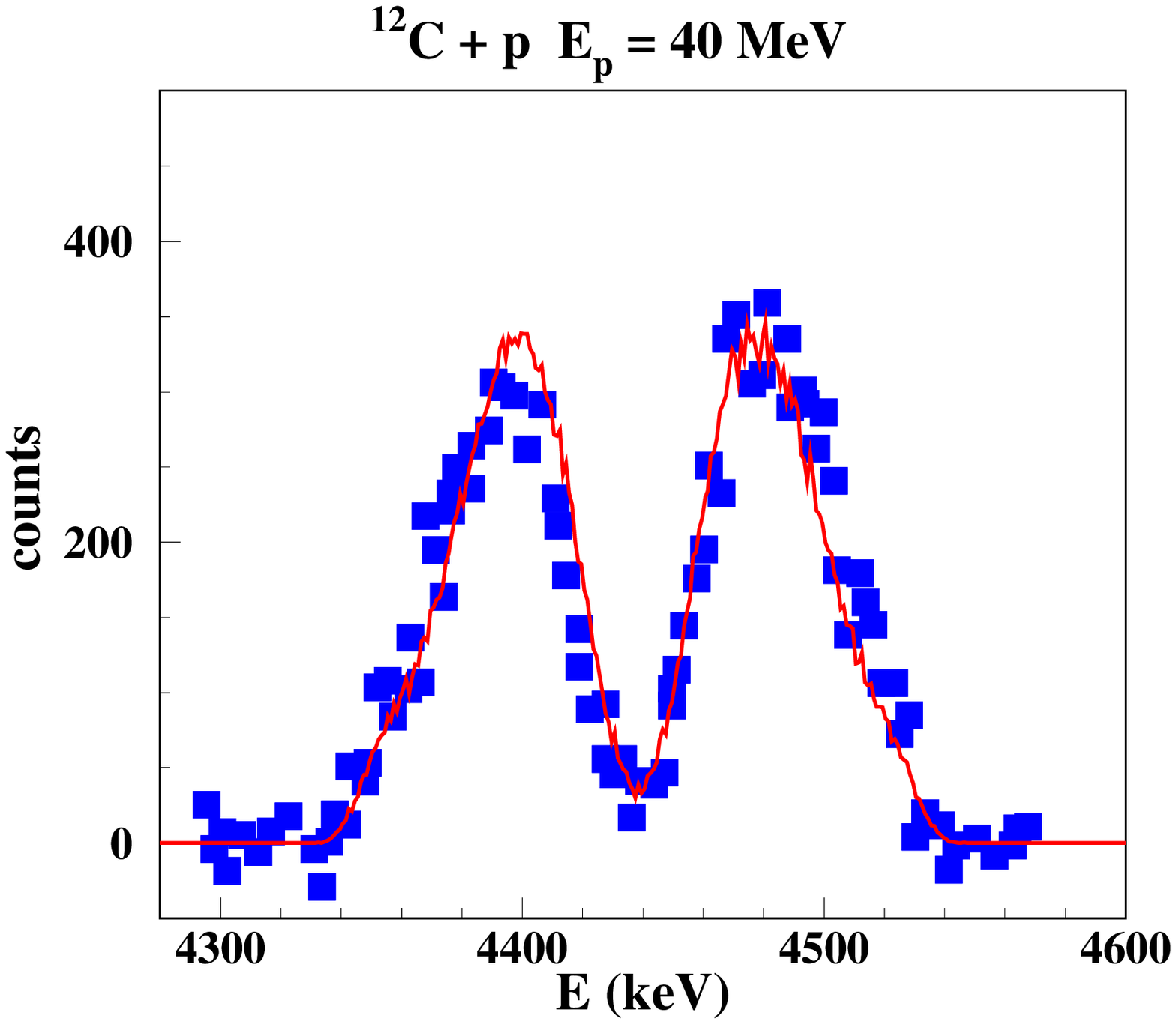}  \includegraphics[width=7.5 cm]{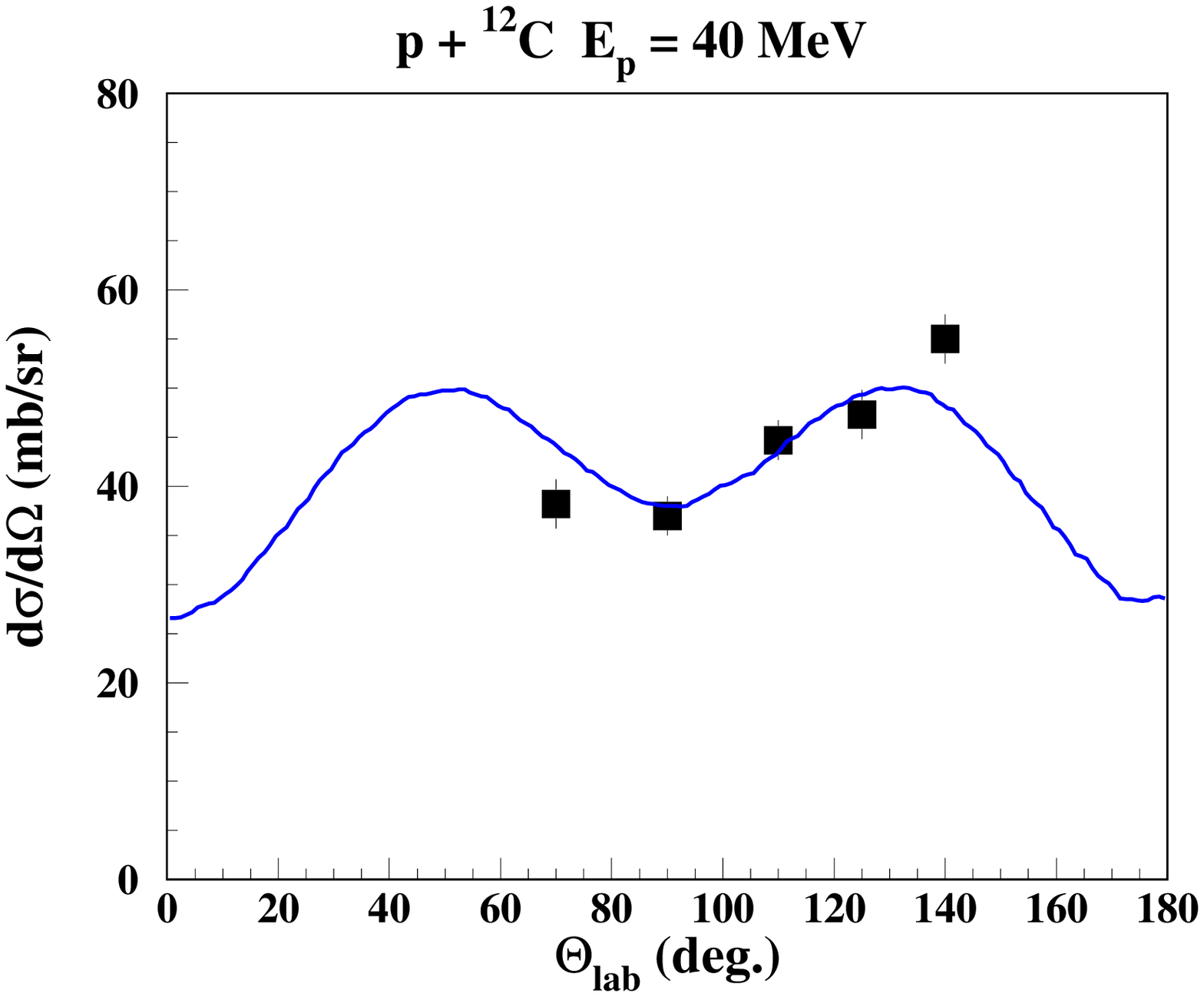}  
\caption{(Color online) Left: Measured shape of the 4.439-MeV $\gamma$-ray line from proton-inelastic scattering off $^{12}$C at $E_p$ = 40 MeV and $\theta$ = 90$^{\circ}$  (Lang et al. \cite{Lang87}, blue squares)  and results of the line shape calculation (red line).  The right panel shows the $\gamma$-ray angular distribution data of Lang et al. (black symbols) and the calculated curve. } 
\label{ShapeL}
\end{figure}

For $E_p$ = 40 MeV, a line shape at $\theta$ = 90$^{\circ}$ and the $\gamma$-ray angular distribution have been published by Lang et al. \cite{Lang87}. The line shape is  well reproduced by a pure direct reaction component, calculated with the potential of Meigooni et al. \cite{Meigooni85}, but there is  an obvious disagreement between the calculated and the mesured angular distribution, see Fig. \ref{ShapeL}. However, the difference of 17\% between the differential cross section data at 70$^{\circ}$ and 110$^{\circ}$  is much larger than the possible assymetry with respect to 90$^{\circ}$ in the laboratory system, having furthermore the wrong sign. This may indicate an underestimation of the error bars, presented to be of the order of 5\%.

Calculations with the nuclear reaction code Talys \cite{Talys} predict a non-negligible contribution of $^{11}$B at $E_p$ = 40 MeV, by deexcitation of its second excited state at 4.445 MeV. Using for the $^{12}$C(p,2p)$^{11}$B reaction the method of Kiener, de S\'er\'eville and Tatischeff \cite{lshape} for the calculation of the 4.439-MeV line in the $^{16}$O(p,p$\alpha$)$^{12}$C reaction, the 4.445-MeV line at 90$^{\circ}$ presents a flat profile in the range 4350 - 4530 keV and an isotropic angular distribution. The deep valley at the nominal line energy of the measured 4.439-MeV line shape (Fig. \ref{ShapeL}) leaves practically no space  for a sizeable $^{11}$B contribution. This contribution was consequently neglected for the line shape calculations in the next section.

\section{Discussion and applications}

Line shapes and $\gamma$-ray angular distributions are well reproduced at all energies up to $E_p$ = 20 MeV with one single $J^{\pi}$ for the CN component, that in many cases can be attributed to a known state in $^{13}$N. There is a strong CN component up to $E_p$ $\sim$12 MeV, above that the direct reaction reaction component dominates. For $E_p$ = 20-25 MeV, only the $\gamma$-ray angular distributions of the Orsay-2002 experiment favor a significant CN component, while the angular distributions of the Orsay-1997 and the Washington experiments, as well as the line shapes of Kolata, Auble \& Galonsky and of both Orsay experiments favor a negligible CN component. The latter is also expected from nuclear reaction theory and it is thus highly probable that the CN component is negligible for the 4.438-MeV line emission above $E_p$ = 20 MeV.

\begin{figure} \includegraphics[width=7.5 cm]{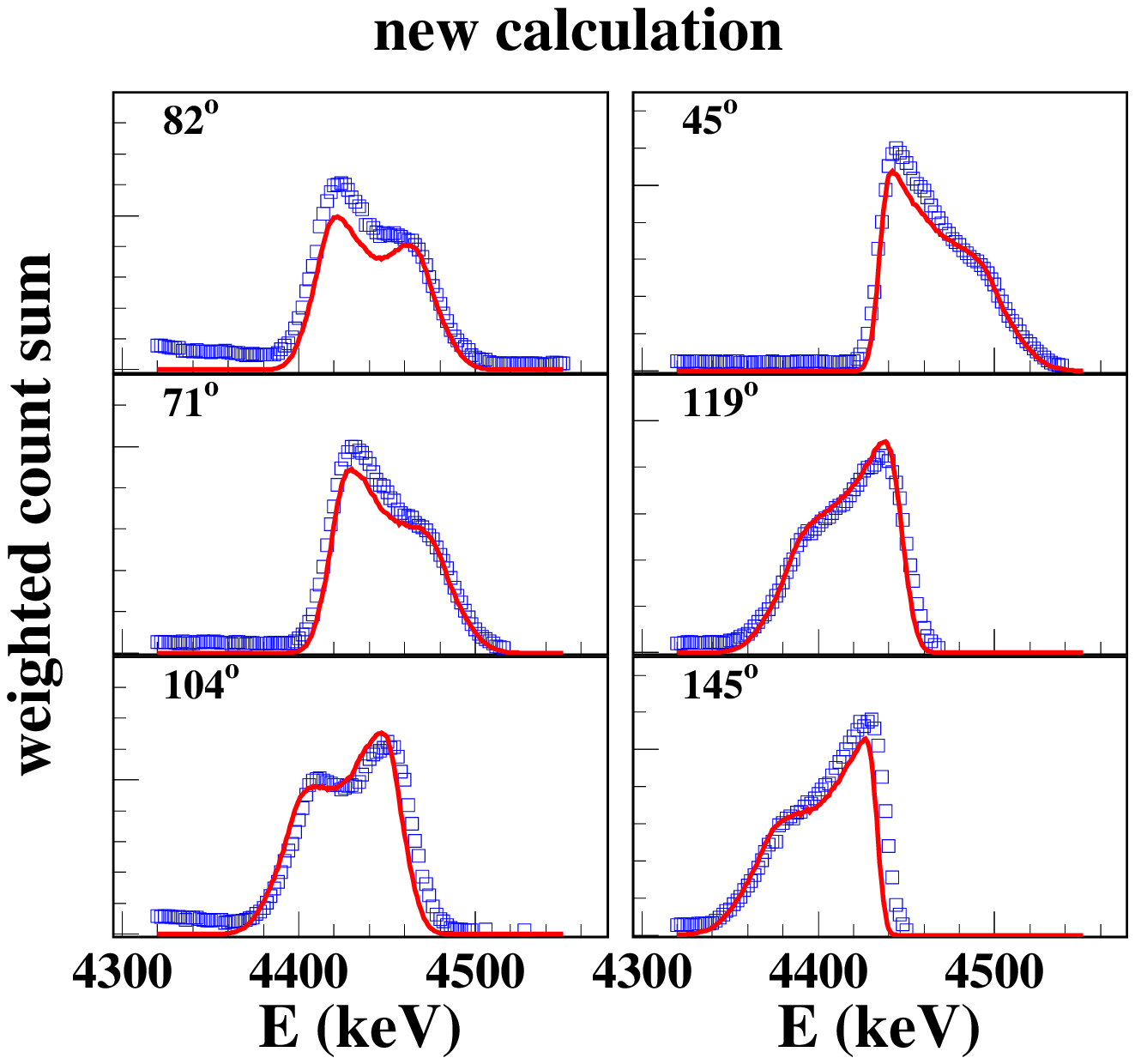}  \includegraphics[width=7.5 cm]{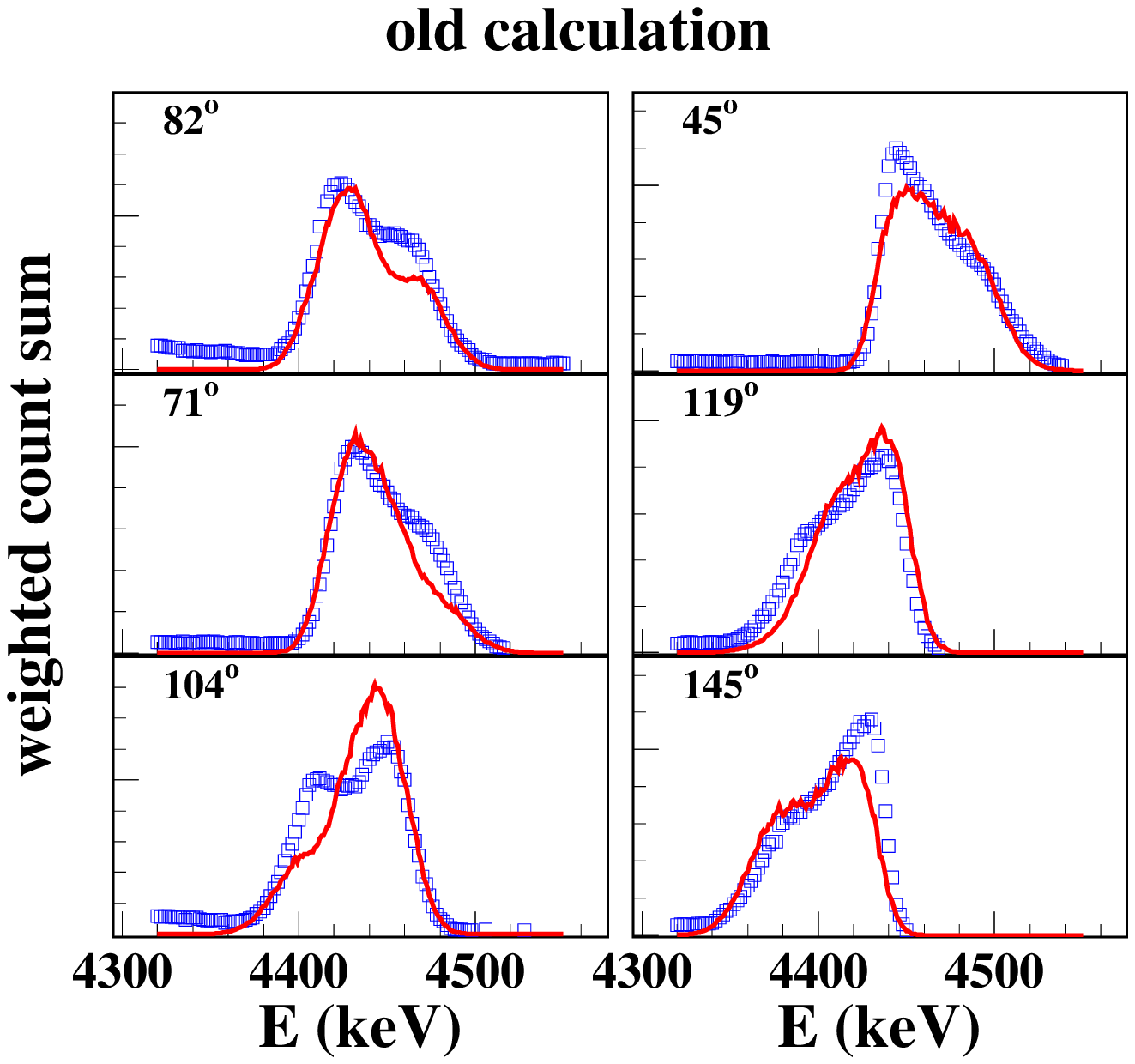}  
\caption{(Color online) Weighted sums of measured shapes of the 4.439-MeV $\gamma$-ray line from proton-inelastic scattering off $^{12}$C in the Orsay-1997 experiment  (blue squares) and calculated line shapes (red lines) from the present study (left) and the previous study (right) of Ref. \cite{lshape}. } 
\label{NewOld}
\end{figure}

Although not all line shapes are perfectly reproduced by the calculations, the present study brings a significant improvement  with respect to previous line shape calculations. The improvement is particularly obvious in the proton energy range with strong CN contribution, where the full treatment of particle-$\gamma$ correlations in the CN component with the angular momentum coupling theory results in much better line shape and angular distribution adjustments than the magnetic-substate population method as proposed in Ref. \cite{RKL} and extensively used in Ref. \cite{lshape}. This is illustrated in Fig. \ref{NewOld} where the line shape data of the Orsay-1997 experiment and calculations are summed with weight factors cor\-responding to a thick-target interaction probability  of an injected beam with energy spectrum $F(E_p) \propto E_p^{-3.5}$, as encountered in solar flares.

At higher energies,  the new data of the Orsay-2002 experiment at $E_p$ = 22.5 and 25 MeV and the data of Refs. \cite{Kolata67,Lang87} at $E_p$ = 23 MeV and 40 MeV, respectively, consolidate the results of Refs. \cite{lshape,Werntz90} pointing out  a largely dominating direct reaction component.  For this component, the potential of Meigooni et al. \cite{Meigooni85} is very successful, predicting correctly the line shapes up to $E_p$ = 40 MeV and the angular distributions up to $E_p$ = 25 MeV with certainty, and very probably up to $E_p$ = 40 MeV. It is worth noting that this potential, developed for elastic and inelastic nucleon scattering off $^{12}$C in the range $E_{n}$ = 20 - 100 MeV should then also give quite reasonable line shapes and $\gamma$-ray angular distributions  up to 100 MeV.
It is thus a very important ingredient for applications requiring calculations in a large proton energy range. With the present results, the 4.438-MeV line emission for solar-flare conditions and in proton radiotherapy of eye tumors can be calculated. In both sites, it is a thick target interaction where proton reactions below $E_p$ = 100 MeV are dominating the line emission.

\subsection{Solar flares}

 $\gamma$-ray emission from the Sun during strong solar flares is regularly observed by orbiting spacecraft since more than 40 years. The high-energy photon spectrum often features some strong, relatively narrow nuclear $\gamma$-ray lines sitting on a quasi-continuum composed  of thousands of other, weaker or broader, nuclear lines and a smooth spectrum from electron bremsstrahlung (see e.g. \cite{SMMcat}). This emission is thought to be produced in impulsive solar flares by energetic particles that are accelerated in the solar corona up to GeV energies and trapped in magnetic loops on the solar surface. The interactions leading to electron bremsstrahlung and nuclear $\gamma$-ray line emissions happen then essentially in the solar chromosphere and photosphere, where the energetic particles are eventually stopped. 

Important questions in solar flares are the composition, the energy spectrum and the directional distribution of the energetic particles, that are linked  to the mechanism that accelerates and propagates them in the solar atmosphere. Composition and directional distribution, and to a lesser extent the energy spectrum, can be deduced from the observed shapes of the strongest nuclear $\gamma$-ray lines. The 4.438-MeV line is often, together with 2.225-MeV neutron capture line on $^1$H and the 511-keV annihilation line, the strongest line from nuclear interactions.  The narrow component of the 4.438-MeV line in solar flares is mainly produced in interactions of energetic protons and $\alpha$ particles with $^{12}$C and $^{16}$O of the solar atmosphere \cite{RKL}. The line is also produced in reverse kinematics by energetic $^{12}$C and  $^{16}$O interacting with ambient $^1$H and $^4$He, but with a very large width and it is hardly detectable.

The line-shape calculations are done with a solar flare loop model of the energetic particle interactions as detailed in Refs. \cite{Hua89,MHKR90}, that describes many features of the observed solar flare $\gamma$-ray emission. Accelerated particles are injected isotropically at the top of a magnetic loop, consisting of an arc in the solar corona connected to two straight portions along a solar radius through the chromosphere down to the loop footpoints in the photosphere, with field strength B constant in the corona and depending on the pressure in the other regions. Particle transport along the magnetic field lines is calculated, including pitch-angle scattering on MHD turbulence until the particle is absorbed in a nuclear interaction or stopped. The interactions of the energetic particles in this model are predominantly induced by downward-directed particles inside the chromosphere. 

For the spectrum and composition of the incident, accelerated particle population the values of Ref. \cite{Kiener06} are used, which were deduced from the analysis of the October 28, 2003 solar flare. This flare was observed by the gamma-ray spectrometer SPI onboard the INTEGRAL satellite \cite{Gros04} and features the most precise data available for the 4.438-MeV line in solar flares. In the described solar flare loop model for this case, slightly more than half  of the 4.438-MeV $\gamma$ rays accounting for the narrow component are produced by proton inelastic scattering off $^{12}$C. Also, more than 90\% of these proton inelastic scattering reactions take place at energies lower than $E_p$ = 25 MeV, completely covered by the calculations, validated by experimental data, of the present study. 

A detailed analysis of the 4.438-MeV $^{12}$C and 6.129-MeV $^{16}$O lines observed from that flare was made in Ref. \cite{Kiener06},  based on the line shape calculations of Ref. \cite{lshape}.  There, both observed lines could be very well reproduced  with relatively narrow downward-directed energetic particle distributions and, for a simultaneous fit of the 4.438-MeV and 6.129-MeV lines, with an $\alpha$-to-proton ratio of $\sim$0.1.  In particular, it could be shown that a narrow particle distribution around the flare axis as resulting from pitch-angle scattering (PAS) with a small mean free path, defined as the PAS($\lambda$=30) distribution in the model of \cite{Hua89,MHKR90}, could perfectly describe both lines, with the flare axis being perpendicular to the solar surface at the flare coordinates. 


Taking the parameters of Ref. \cite{Kiener06} for the carbon-to-oxygen ratio in the solar atmosphere and the energy spectra of the incident particles, the 4.438-MeV line shape in the present study was calculated for the PAS($\lambda$=30) directional distribution and a range of energetic $\alpha$-to-proton ratios ($\alpha/p$). The calculation of proton and $\alpha$-particle reactions with $^{16}$O was done following the method described in \cite{lshape}. For the $^{16}$O(p,p$\alpha$)$^{12}$C reaction, isotropic emission of proton, $\alpha$ particle and  $\gamma$ ray perfectly reproduced measured angular distributions and line shapes below $E_p$ = 20 MeV. Extrapolation to higher energies and to the $^{16}$O($\alpha$,2$\alpha$)$^{12}$C reaction should be quite accurate in this case. 

\begin{figure} \includegraphics[width= 7.5 cm]{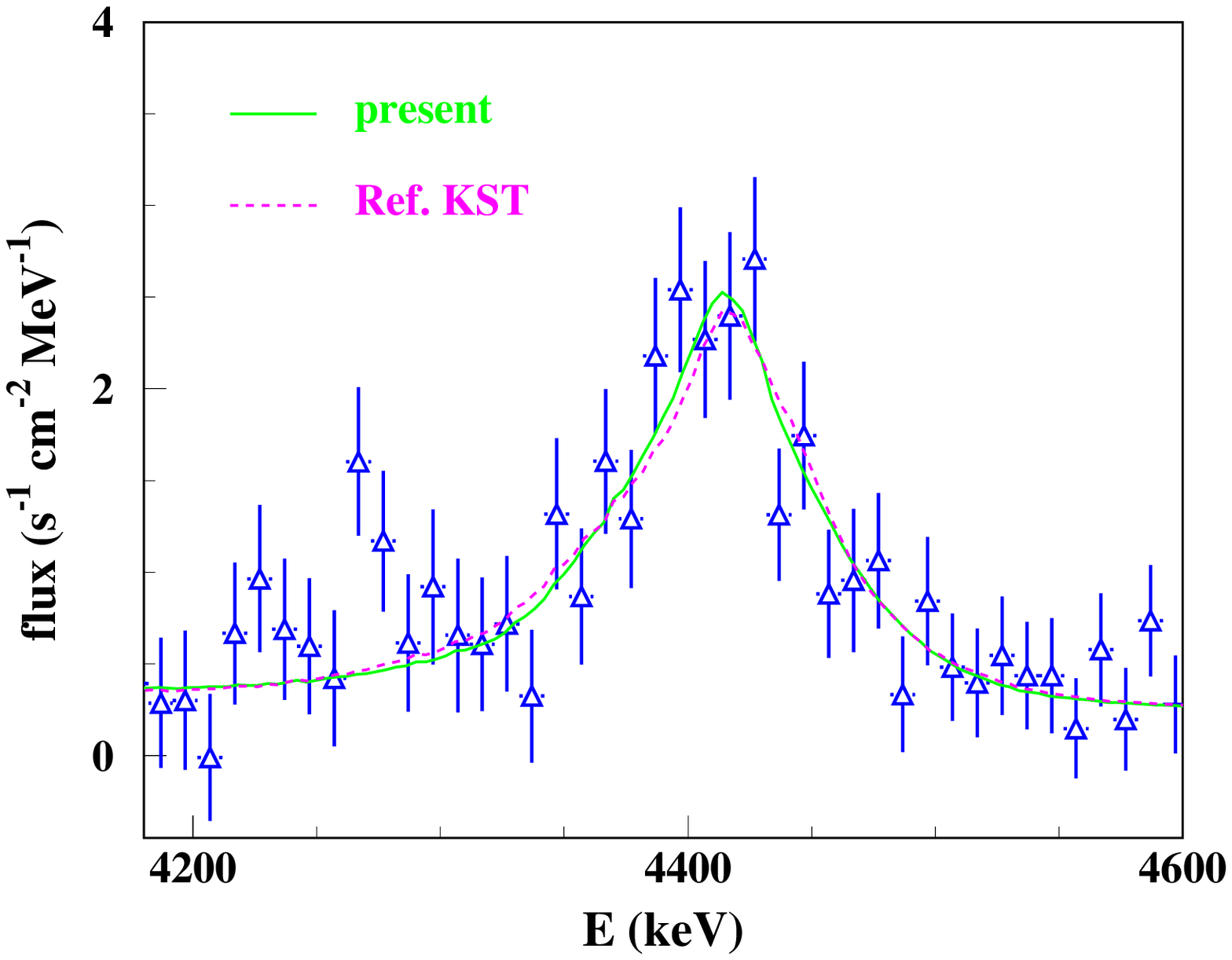} \includegraphics[width=7.5 cm]{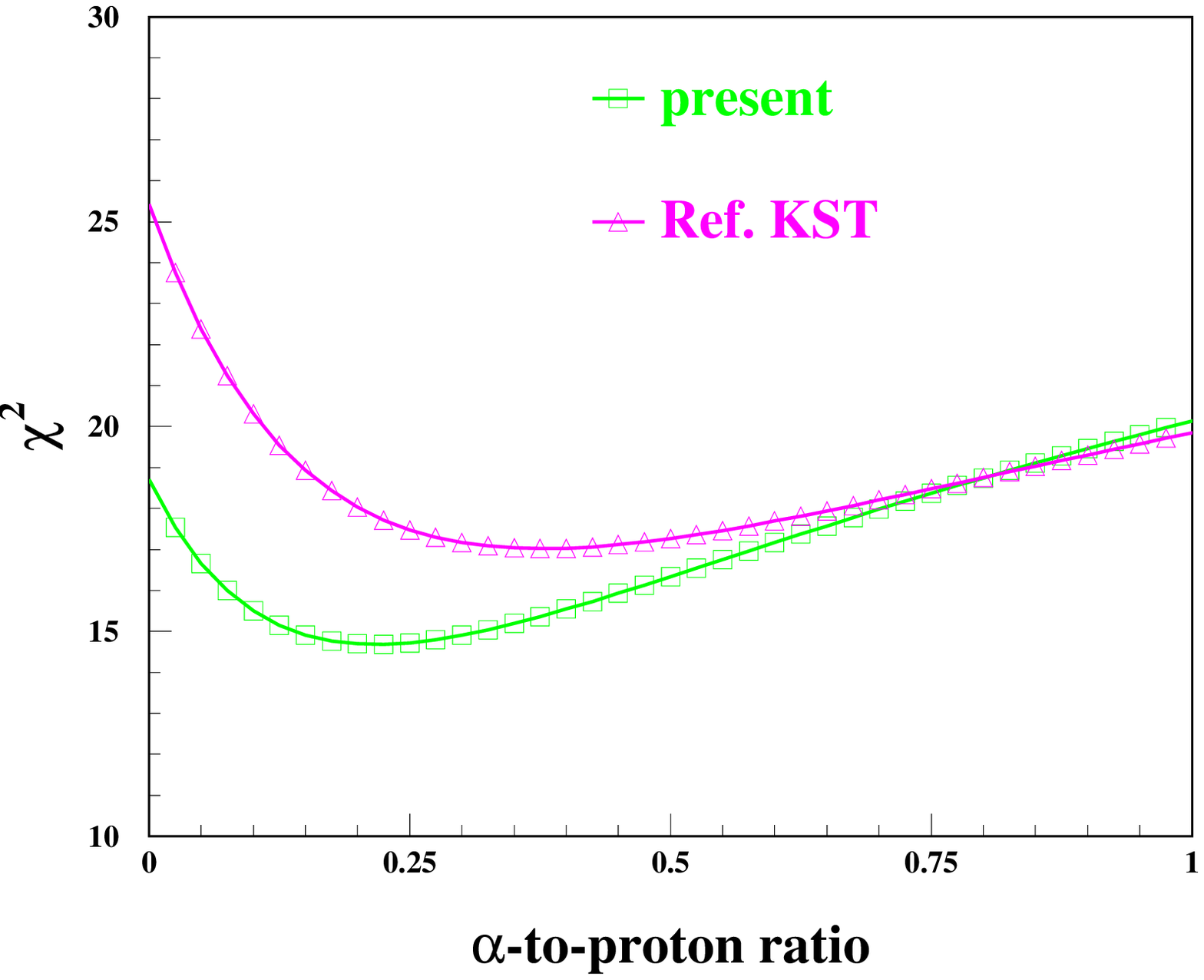} 
\caption{ (Color online) Left: Symbols show the observed spectrum around the 4.438-MeV line from the Oct. 28, 2003 solar flare, with Compton and continuum backgrounds subtracted (see Ref. \cite{Kiener06} for details). The full green and dashed magenta lines represent the best-fit results obtained with present line-shape calculations  ($\alpha/p$ = 0.22$^{+0.20}_{-0.13}$) and with previous line-shape calculations ($\alpha/p$ = 0.38$^{+0.30}_{-0.18}$) (Ref.  KST \cite{lshape}), respectively. The results of $\chi^2$ minimization in the spectrum energy range $E$ = 4300-4580 keV as a function of $\alpha/p$ is shown on the right panel.}
\label{Cline}
\end{figure}

For proton inelastic scattering off $^{12}$C, the previous line shape calculations \cite{lshape} and the present ones have been used. The best fit $\alpha/p$  and the resulting line shapes, compared with the observed data, are shown in Fig. \ref{Cline}. A nearly perfect fit could be obtained with both line shape calculations with, however, significantly different $\alpha/p$. This result is in line with the findings of  Ref. \cite{Kiener06} about the directional distribution of the interacting energetic particles, but shows also the sensitivity to small differences in the line shape calculations for inelastic proton scattering. It illustrates clearly the need for accurate line-shape calculations in a wide particle energy range, as it is done for the first time in the present work. 

It is worth mentioning also that in this example, $\alpha$-particle inelastic scattering plays a non-negligible role. Its contribution to the 4.438-MeV emission is for example about 30\% for  $\alpha/p$ = 0.22.  Still higher $\alpha/p$ values of $\sim$0.5 have been proposed to prevail in impulsive solar flares \cite{MRKR}. For the future, an improvement in the line shape calculations for $\alpha$-particle inelastic scattering  would therefore be very appreciable for such studies. 

\subsection{Proton radiotherapy}

\begin{figure} \includegraphics[width= 7.5 cm]{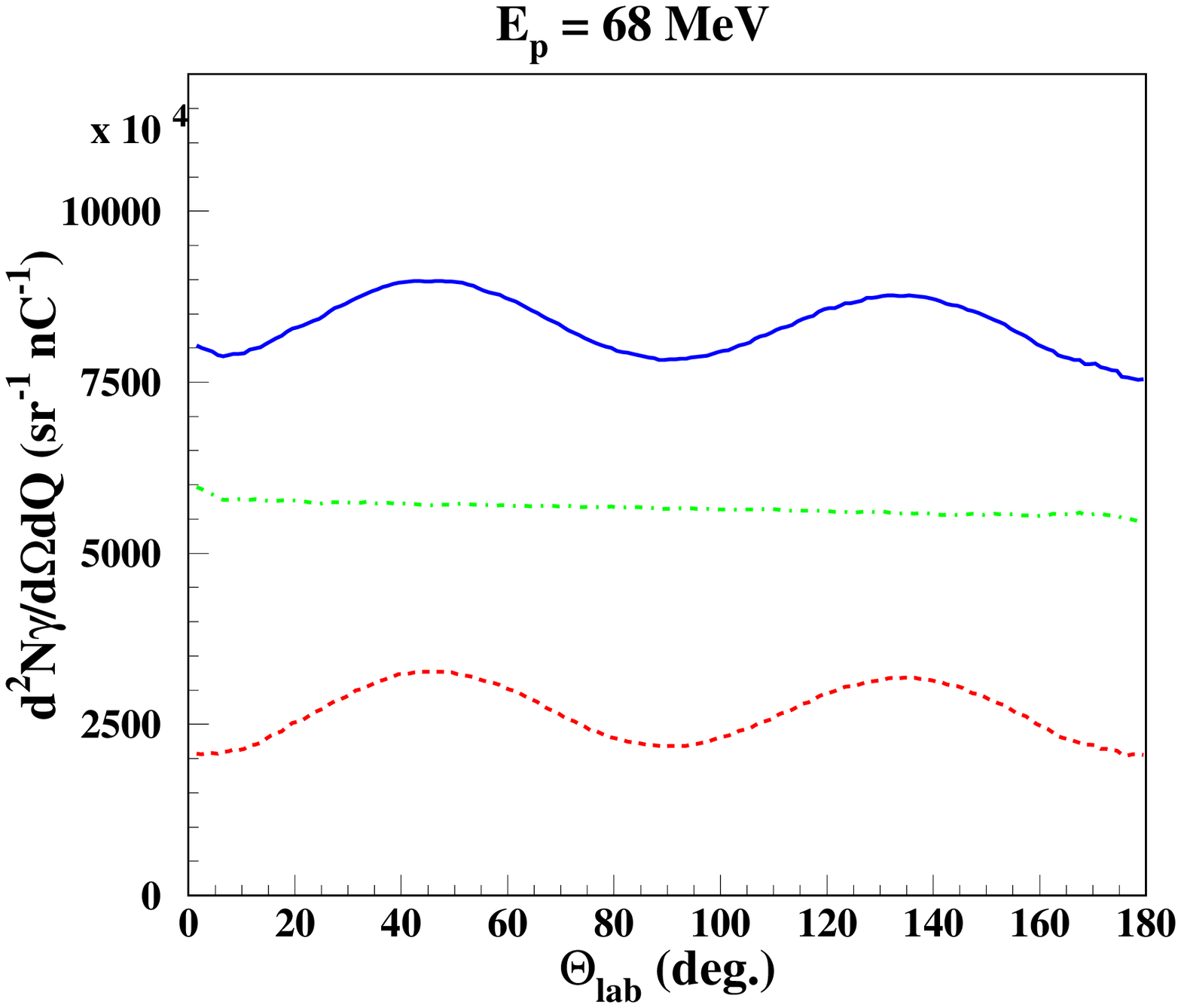} \includegraphics[width=7.5 cm]{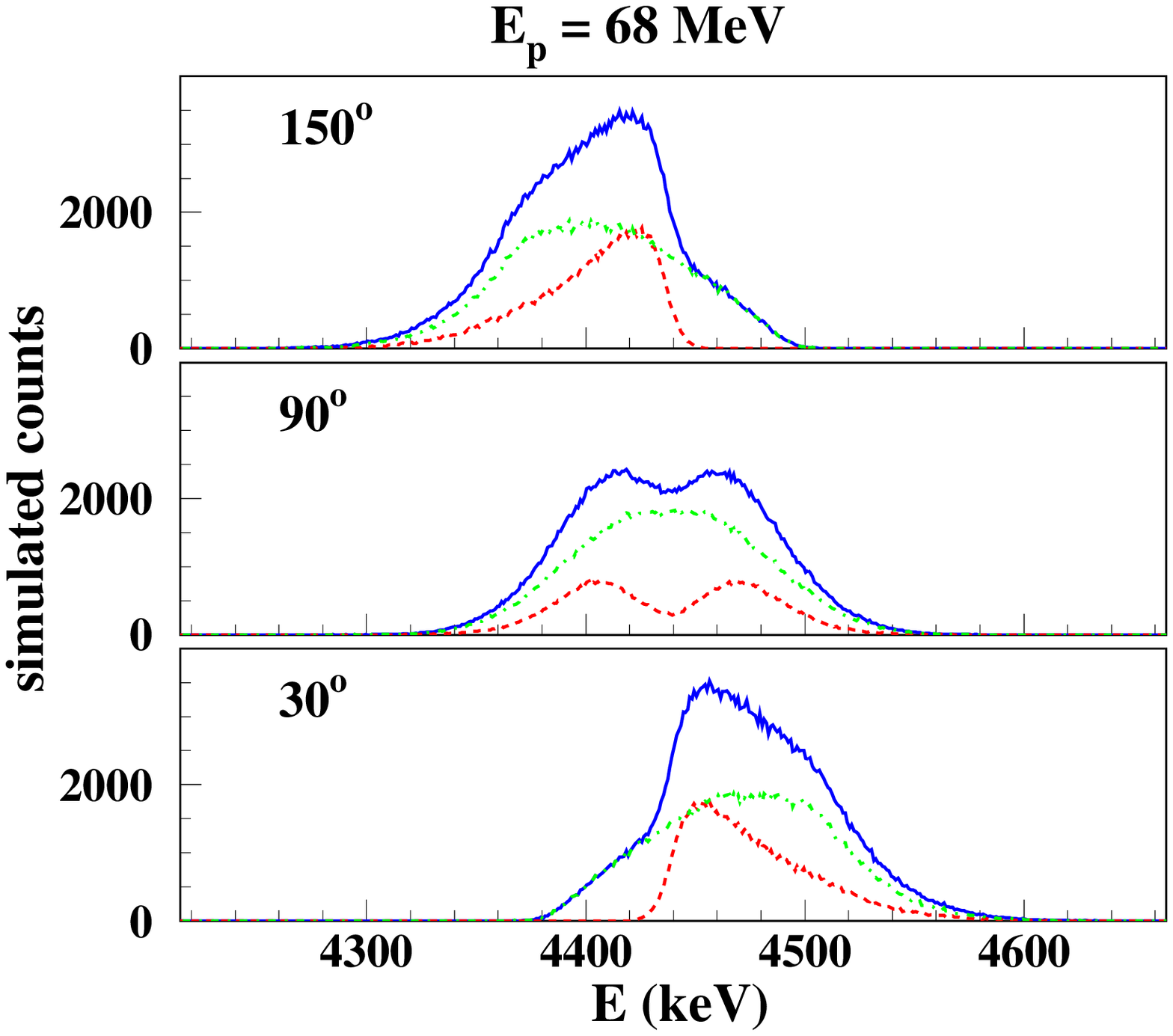} 
\caption{ (Color online) Angular distribution (left panel) and line shapes (right panel) of the 4.439-MeV $\gamma$-ray line emitted in reactions of a 68-MeV proton beam stopped in human tissue.  Solid blue lines represent the sum of proton reactions with $^{12}$C and $^{16}$O, dashed red lines show the component of reactions with $^{12}$C and dot-dashed green lines represent the contribution of reactions with $^{16}$O.}
\label{DistPT}
\end{figure}

Similar to solar flares, the 4.438-MeV line in human tissue during proton radiotherapy is essentially produced by inelastic scattering off $^{12}$C and reactions with $^{16}$O. Incident proton energies are in the range of about 60 MeV to 200 MeV. For an incident proton energy of $E_p$ = 68 MeV, typical for eye cancer treatment, there are 2.7$\cdot$10$^{-3}$  4.438-MeV $\gamma$ rays  emitted per incident proton due to inelastic scattering off $^{12}$C, and  5.6$\cdot$10$^{-3}$ per incident proton due to proton reactions with $^{16}$O and a negligible part ($\sim$10$^{-4}$ per incident proton) is from reactions with $^{14}$N. The calculated angular distribution and shapes of the 4.438-MeV line  are shown on Fig. \ref{DistPT}. The calculations were done with the average composition of the human body \cite{Emsley}, the experimental cross sections  for the 4.438-MeV  line production in proton inelastic scattering off $^{12}$C, the cross sections of table I in Ref. \cite{lshape} for  proton reactions with $^{16}$O and from the cross section compilation of Ref. \cite{MKKS} for proton reactions with $^{14}$N. The 4.438-MeV line yield from proton reactions with $^{16}$O  is reduced by about 25\% if one uses  the compilation, Ref.\cite{MKKS} for cross sections with $^{16}$O instead of Ref. \cite{lshape}. The energy loss of protons in human tissue is calculated with the tables of SRIM \cite{SRIM}.

The resulting 4.438-MeV line exhibits characteristic  shapes at the different observation angles. The knowledge of these shapes is essential for the extraction of the 4.438-MeV line integral in $\gamma$-ray spectra eventually measured by radiation monitor detectors.  Inelastic scattering modulates the $\gamma$-ray angular distribution, with a difference of about 10-15\% between the maximum at 45$^{\circ}$ and the minima.  At this incident proton energy, slightly more than half of the $\gamma$-ray emission in reactions with $^{12}$C is produced for proton energies above $E_p$ = 25 MeV, and about 80\% in reactions with $^{16}$O. The  calculations are thus strongly based on extrapolations, but line shapes and angular distributions should nonetheless be quite accurate, as explained above. The present study should therefore allow a more precise monitoring of the radiation dose, independent of the position of the radiation monitor detector with respect to the beam direction. This may not hold for proton energies well above $E_p$ = 100 MeV, where other potentials may be needed for the inelastic scattering off $^{12}$C. More importantly, cross sections of the 4.438-MeV line emission in proton reactions with  $^{16}$O have only been measured up to $E_p$ = 50 MeV, and there is a strong discrepancy at $E_p$ = 40 MeV between the data of Refs. \cite{Lang87} and \cite{Lesko88}. 

\section{Summary}

In this paper, a new approach leading to a detailed and accurate calculation of the 4.438-MeV $\gamma$-ray emission in proton inelastic scattering off $^{12}$C has been presented. It is based on results of coupled-channels calculations for the direct reaction component and angular momentum coupling theory for CN resonances. Parameters of the calculations were adjusted by comparison with a comprehensive data set of measured line shapes and $\gamma$-ray angular distributions in the proton energy range from threshold to $E_p$ = 25 MeV. All experimental data could be relatively well reproduced by a set of isolated CN resonances of definite spin and parity incoherently added to the direct reaction component.  With this study, a substantial improvement with respect to previous line shape calculations, in particular in the energy range of dominating CN resonances could be obtained.

At energies above $E_p$ $\sim$14 MeV, where the direct reaction dominates, the data are well described by the results of coupled-channels calculations with an energy- and channel-dependent potential, that has been derived for nucleon scattering off $^{12}$C in the range up to $E_p$ = 100 MeV. This enables straightforward extrapolations to higher proton energies, and the scarce data available in this range, one line shape and the $\gamma$-ray angular distribution at $E_p$ = 40 MeV, are also reasonably described by these calculations.  With the present results, line shapes and $\gamma$-ray angular distributions can be predicted with accuracy for proton inelastic scattering off $^{12}$C from threshold to at least $E_p$ = 100 MeV. 

Applications to solar-flare $\gamma$-ray emission and proton radiotherapy were discussed, and differences with a previous method outlined.  The 4.438-MeV $\gamma$-ray emission in solar flares can now be accurately calculated for not too high $\alpha$-to-proton ratios, but would greatly benefit from an improvement in the treatment of inelastic $\alpha$-particle scattering. Applications to proton radiotherapy would  also benefit from new experimental data, in particular for proton reactions with $^{12}$C above $E_p$ = 100 MeV and $^{16}$O above $E_p$ = 30 MeV. Fortunately, new experimental data for p + $^{12}$C and p + $^{16}$O reactions up to $E_p$ = 200 MeV from the iThemba LABS cyclotron should soon be available. First results are encouraging \cite{Yahia}, such that still more accurate predictions in the cited applications can be possible in the near future.

\begin{acknowledgments} 
I like to express my gratitude to I. Deloncle and V. Tatischeff  for their suggestions and critical reading of the manuscript.
\end{acknowledgments}

\section{Appendix}

\subsection{Compound-nucleus resonances}

In equation \ref{W}, the efficiency tensor $\epsilon^{\star}_{k_{L} q_{L}}$  of the emitted proton in the CN resonance decay for a counter with axial symmetry, can be derived from equations (2.52) and (2.78) in Ferguson \cite{Ferguson}.  In the present calculations,  the scattering plane is taken as the y - z plane of the laboratory system, thus $\phi_p$ = 0, and ideal counters for all particles are supposed. For a beam of unpolarized protons:

\begin{equation} 
\epsilon_{kq}(L L') ~ = ~ (-1)^{s + k - L} ~ \frac{\hat{l}^2 \hat{L} \hat{L}'}{ \hat{k} \sqrt{4\pi}} ~ < l 0 l 0 \mid k 0 > ~ W(l l L L'; k s)~  ~ Y_{kq}^{\star}(\theta_p,0) 
\label{eLL}
\end{equation}

where $L$ results from the vector sum of proton spin $s$ and orbital angular momentum $l$.  

The efficiency tensor for the $\gamma$-ray counter $\epsilon^{\star}_{k_L{_{\gamma}} q_L{_{\gamma}}}$, with $L_{\gamma}$ =  $L'_{\gamma}$ = $c$  for the $2^+$ $\rightarrow$ $0^+$ transition and no polarization and parity mixture (eqs. (2.57) and (2.78) in Ferguson), is:

\begin{equation} 
\epsilon_{kq}(L_{\gamma} L'_{\gamma}) ~ = ~ (-1)^{c - 1} ~  \frac{\hat{c}^2}{ \hat{k} \sqrt{4\pi}} ~ < c 1 c -1 \mid k 0 > ~ Y_{kq}^{\star}(\theta_{\gamma},\phi_{\gamma})
\end{equation}

The term $f_c$ with angular couplings is:

\begin{equation} 
f_c ~ = ~ \hat{b}^2 \hat{k_c} \hat{k_{L}} ~ < k_c q_c k_{L} q_{L} \mid k q > ~ W_{9}(c L b; c L' b; k_c k_L k)
\end{equation}

where $W_{9}$ means the Wigner 9-j coefficient with indices written line by line.

The probability for inelastic scattering with scattering angle  $\theta_p$ can be obtained from equation \ref{W} with the efficiency tensor  of unobserved $\gamma$ ray and excited nucleus 

\begin{equation}
\epsilon_{kq}(L_{\gamma} L'_{\gamma}) = \hat{c} ~ \delta_{k_c 0} \delta_{q_c 0} \delta_{c c'}
\end{equation}

and integration over $\theta_{\gamma}$ and $\phi_{\gamma}$: 

\begin{equation}
W(\theta_p) = \sum_{}^{} t_{kq}(b) ~ \epsilon^{\star}_{kq}(L L') ~ \hat{b}^2 (-1)^{c+b+k+L} ~ W(b L b L';ck) M(cLb) M(cL'b)^{\star}
\end{equation}

with $\epsilon_{kq}(L L')$ of equation \ref{eLL}, and the sum running over $k$, $q$, $L$ and $L'$. The differential cross section $d\sigma/d\Omega (\theta_p)$ is proportional to $W(\theta_p)$.

\subsection{Direct reactions}

In equation \ref{Wdir},  Satchler's gamma-radiation parameters (eqs. (10.153) (10.154) in \cite{Satchler}) are, with $\mid g_l \mid$ =  1; $k$, $l, l'$ = even and $I_c$ = 0:

\begin{equation}
R_k (\gamma) = \hat{b} \hat{l}^2 ~ (-1)^{b+1} ~ < l 1 l -1 \mid k 0 > ~ W( l l b b; k 0 )
\end{equation}

When using the nuclear reaction code Ecis97  \cite{Ecis} for OM calculations, the scattering amplitudes $T_{\beta,\sigma_o,\alpha,\sigma_i}(\theta_p)$ have to be constructed from other output, like the scattering matrix elements $S^j_{l_o j_o l_i j_i}$ (see equation 16).  As this construction happened to be among the more laborious tasks in the calculations, an outline is therefore given here.  In the spin-orbit coupling scheme with proton spin $s$, and $i(o)$ for incoming (outgoing) channel:

\begin{equation}   \vec{l}_{i(o)} + \vec{s}_{i(o)} = \vec{j}_{i(o)} ~ ~ ~ ~ ~ \vec{j}_{i(o)} + \vec{a}(\vec{b}) = \vec{j} 
\end{equation}

they are given by eq. (5.45) in Satchler, in the scattering plane ($\phi_p$ = 0) with the incoming beam in direction of the z-axis ($\lambda_i$ = 0) and $a$ = 0:

\begin{eqnarray} 
T_{\beta,\sigma_o,\alpha,\sigma_i}(\theta_p) ~ = ~ \frac{2\pi}{k_{\alpha}} ~ \sum ~ \frac{\hat{l}_i}{\sqrt{4\pi}} ~ <l_i s_i 0 \sigma_i \mid j_i   \sigma_i> ~ < j_i 0 \sigma_i 0 \mid j \eta> ~  \nonumber \\  < l_o s_{o} \lambda_o \sigma_{o}  \mid j_o (\lambda_o+\sigma_{o}) >   < j_o b (\lambda_o+\sigma_o) \beta \mid j \eta > \nonumber \\  e^{i \sigma_{ps}} ~ (S^j_{l_o j_o l_i j_i} - \delta_{l_i l_o} \delta_{j_i j_o}) ~ Y_{l_o,\lambda_o}(\theta_p,0)
\end{eqnarray}

where $\lambda,\alpha,\beta,\sigma,\eta$ are the spin projections of orbital angular momenta $l$, target ground state spin $a$ and excited state spin $b$, proton spin $s$  and coupled spins $j$, respectively. $\sigma_{ps}$ is the phase (in Ecis97 output: phase with Coulomb) of the scattering matrix $S$. The sum runs over $l_i$, $l_o$, $j_i$ and $j_o$;  here $j$ = $j_i$  for $a$ = 0.

\bibliography{mybib}

\newpage

\begin{table} 
\caption{Results of line shape and $\gamma$-ray angular distribution calculations for the Orsay experiment in 1997. $E_p$ and $\Delta E_p$ are the beam energy and energy loss in the target in MeV, respectively, and $E_x$ the excitation energy in $^{13}$N corresponding to $E_p$-$\Delta E_p$/2.  $J^{\pi}$ and $W_{CN}$ are the spin-parity and proportion of the CN component of the best adjustment.  The probable corresponding state in $^{13}$N, with its excitation energy in MeV (values from \cite{NNDC}), is shown in column 9. $\chi^2$(C) is the result of the least-squares fit of the measured data with the calculated angular distributions and $\chi^2$(L) the result of the corresponding Legendre-polynomial fit. Values in italics for $W_{l_0}$, the proportion of the lowest possible decay angular momentum,  are not constraint.}
\begin{tabular}{lllllllllll}
 $E_p$ ~ &  $\Delta E_p$ ~ & $E_x$ ~ &  $W_{CN}$  &  $J^{\pi}$  ~ & $W_{l_0}$~ & ~ $\chi^2$(C) & ~ $\chi^2$(L) &~ $^{13}$N state & ~~ remarks \\
\hline 
8.6 & 0.16 & 9.8 & 0.35 & $\frac{1}{2}^+$ & 1.0  & ~ 23 & ~ 9.5 & ~ ($\frac{1}{2}^+$), 10.25(15) & \\
9.0 & 0.15 & 10.2 & 0.35 & $\frac{1}{2}^+$ & 1.0     & ~ 9.2 & ~ 6.1 & ~ ($\frac{1}{2}^+$), 10.25(15) & \\
9.2 & 0.15 & 10.4 & 0.66 & $\frac{1}{2}^+$ & 1.0   & ~ 5.3 & ~ 7.0 & ~ ($\frac{1}{2}^+$), 10.25(15) & ~ no line shapes \\
9.2 & 0.15 & 10.4 & 0.78 & $\frac{7}{2}^-$ & 0.7   & ~ 5.1 & ~ 7.0 & ~ $\frac{7}{2}^-$, 10.360 & ~ no line shapes \\
9.6 & 0.15 & 10.7 & 0.53 & $\frac{1}{2}^-$ & {\it 0.5}     & ~ 3.4 & ~ 3.8 & ~ $\frac{1}{2}^-$, 10.833(9) &  \\
10.0 & 0.14 & 11.1 & 0.49 & $\frac{5}{2}^+$ & 0.3   & ~ 3.0 &  ~ 3.5 & ~ $\frac{5}{2}^+$, 11.530(12) & \\
10.6 & 0.14 & 11.7 & 0.76 & $\frac{3}{2}^+$ & 0.2     & ~ 2.5 & ~ 3.1 & ~ $\frac{3}{2}^+$, 11.740(40) &  \\
11.0 & 0.13 & 12.0 & 1.00 & $\frac{7}{2}^-$ & 0.8     & ~ 7.0 & ~ 6.4 & ~ $\frac{7}{2}^-$, 12.130(50) & \\
11.4 & 0.13 & 12.4 & 0.95 & $\frac{7}{2}^-$ & 0.8     & ~ 6.7 & ~ 7.9 & ~ $\frac{7}{2}^-$, 12.130(50) & \\
12.0 & 0.12 & 13.0 & 0.34 & $\frac{7}{2}^-$ & 0.4     & ~ 10.4 & ~ 10.2 &  ~ ( ), 12.937(24) & \\
12.6 & 0.02 & 13.6 & 0.08 & $\frac{3}{2}^+$ &  0.0     & ~ 13.6 &  ~ 10.6 & ~ $\frac{3}{2}^+$, 13.50(20) & \\
13.0 & 0.02 & 13.9 & 0.12 & $\frac{3}{2}^+$ & 0.2      & ~ 17.8 & ~ 14.2  & ~ $\frac{3}{2}^+$, 14.050(20) & \\
13.6 & 0.02 & 14.5 & 0.21 & $\frac{3}{2}^+$ & 0.4     & ~ 13.4 &  ~ 10.2 & ~ $\frac{3}{2}^+$, 13.50(20) & \\
14.0 & 0.02 & 14.8 & 0.33 & $\frac{3}{2}^+$ & 1.0  & ~ 14.7 &  ~ 13.3 & ~ ($\frac{3}{2}^+$), 15.30(20) & \\
14.4 & 0.11 & 15.2 & 0.17 & $\frac{3}{2}^+$ & 0.0  & ~ 14.7 &  ~ 11.2 & ~ ($\frac{3}{2}^+$), 15.30(20) & \\
15.2 & 0.10 & 15.9 & 0.00 & $\frac{7}{2}^+$ & {\it 0.5}  & ~ 13.8 & ~ 8.9 & ~ $\frac{7}{2}^+$, 15.990(30) & ~ $\sim$15\% of $^{16}$O(p,p$\alpha$) \\
16.25 & 0.01 & 16.9 & 0.13 & $\frac{5}{2}^+$ &  1.0     & ~ 13.8 & ~ 7.6 &  & \\
17.25 & 0.01 & 17.9 & 0.20 & $\frac{3}{2}^+$ & 1.0      & ~ 8.2 & ~ 5.1 & ~ $\frac{3}{2}^+$, 18.150(30) & \\
18.25 & 0.01 & 18.8 & 0.10 & $\frac{5}{2}^-$  & 1.0      & ~ 10.9 & ~ 6.8 & ~ $\frac{5}{2}^-$, 19.830 &  \\
19.75 & 0.01 & 20.2 & 0.10 & $\frac{5}{2}^-$ & 1.0      & ~ 11.8 & ~ 5.8 & ~ $\frac{5}{2}^-$, 19.830  \\
\hline  
 \end{tabular}
\label{tab97}

\end{table}

\begin{table} 
\caption{Results of line shape and $\gamma$-ray angular distribution calculations for the Orsay experiment in 2002. The different columns are explained in table \ref{tab97}. The last column gives the reference of the used optical potential. When no potential is indicated, the one of Meigooni et al. \cite{Meigooni85} was used for the direct component.}
\begin{tabular}{lllllllllll}
 $E_p$ ~ &  $\Delta E_p$ ~ & $E_x$ ~ &  $W_{CN}$  &  $J^{\pi}$  ~ & $W_{l_0}$~ & ~ $\chi^2$(C) & ~ $\chi^2$(L) &~ $^{13}$N state & ~~ remarks \\
\hline 
5.44 & 0.04 & 6.9 & 0.76 & $\frac{3}{2}^+$ & 1.0  & ~ 0.3 & ~ 0.09 & ~ $\frac{3}{2}^+$, 6.886(8) &  \\
5.95 & 0.04 & 7.4 & 0.36 & $\frac{5}{2}^-$  & 0.9 & ~ 0.7 & ~ 0.37 & ~ $\frac{5}{2}^-$, 7.376(9) & pot. PEA72 \cite{PEA72}  \\
6.5 & 0.04 & 7.9 & 0.73 & $\frac{3}{2}^+$ & 0.4  &  ~ 0.4 & ~ 0.002 & ~ $\frac{3}{2}^+$, 7.900 &  \\
7.0 & 0.04 & 8.4 & 0.57 & $\frac{3}{2}^+$ & 0.2 & ~ 0.4 & ~ 0.01 & ~ $\frac{3}{2}^+$, 7.900 & pot. GUR69   \cite{GUR69}  \\
7.6 & 0.04 & 8.9 &  0.61 & $\frac{1}{2}^-$ & {\it 0.5} & ~ 3.5 & ~ 0.46 & ~ $\frac{1}{2}^-$, 8.918(11) & pot. PEA72  \cite{PEA72}   \\
8.0 & 0.04 & 9.3 & 0.44 & $\frac{3}{2}^+$ & 0.6 & ~ 0.3  & ~ 0.18 ~ & ~ $\frac{3}{2}^+$, 7.900 & \\
8.23 & 0.04 & 9.5 & 0.57 & $\frac{3}{2}^+$ & 0.6 & ~ 2.7 & ~ 0.13 &  ~ $\frac{3}{2}^+$, 7.900 & pot. PEA72  \cite{PEA72}  \\
8.4 & 0.04 & 9.7 & 0.45 & $\frac{3}{2}^+$ & 0.7 & ~ 1.3 & ~ 0.02  & ~ $\frac{3}{2}^+$, 7.900 & \\
20.0 & 0.02 & 20.4 & 0.15 &  $\frac{5}{2}^-$  & 0.4 & ~ 3.0 & ~ 0.13  & ~ $\frac{5}{2}^-$, 20.200  &  \\
22.5 & 0.02 & 22.7 & 0.15 &  $\frac{1}{2}^+$  & 1.0 & ~ 2.4 & ~ 0.01  &  ~ $\frac{1}{2}^+$, 22.4(5) &\\
25.0 & 0.02 & 25.0 & 0.15 &  $\frac{1}{2}^+$  & 1.0  & ~ 3.1 & ~ 1.5 & ~  &  \\
\hline  
 \end{tabular}
\label{tab02}

\end{table}

\newpage

\end{document}